\documentclass[useAMS,usenatbib,usegraphicx]{mn2e}
\usepackage{color}

\usepackage[colorlinks,citecolor=black,linkcolor=black,urlcolor=black]{hyperref}
\newcommand*{\figref}[1]{Figure~\ref{#1}}

\newcommand*{\tabref}[1]{Table~\ref{#1}}
\newcommand*{\tabrefs}[1]{Tables~\ref{#1}}
\newcommand*{\eqnref}[1]{Equation~\ref{#1}}
\newcommand*{\eqnrefs}[1]{Equations~\ref{#1}}
\newcommand*{\secref}[1]{Section~\ref{#1}}
\newcommand*{\secrefs}[1]{Sections~\ref{#1}}
\newcommand*{\tablespace}{\vspace{4ex}}

\newcommand*{\unit}[1]{\ensuremath{\mathrm{\, #1}}}
\newcommand*{\erg}{\unit{erg}}
\newcommand*{\keV}{\unit{keV}}
\newcommand*{\second}{\unit{s}}

\newcommand*{\cm}{\unit{cm}}
\newcommand*{\km}{\unit{km}}
\newcommand*{\Mpc}{\unit{Mpc}}
\newcommand*{\kpc}{\unit{kpc}}

\newcommand*{\E}[1]{\ensuremath{\times 10^{#1}}}
\newcommand*{\logTen}{\ensuremath{\log_{10}}}
\newcommand*{\expectation}[1]{\ensuremath{\left\langle #1 \right\rangle}}
\newcommand*{\like}{\ensuremath{\mathcal{L}}}
\newcommand{\degrees}{\ensuremath{^{\circ}}}

\newcommand*{\Msun}{\ensuremath{\, M_{\odot}}}
\newcommand*{\mysub}[2]{\ensuremath{#1_{\mathrm{#2}}}}

\newcommand*{\Omegam}{\mysub{\Omega}{m}}
\newcommand*{\Omegab}{\mysub{\Omega}{b}}

\newcommand*{\LCDM}{\ensuremath{\Lambda}CDM}
\newcommand*{\fgas}{\mysub{f}{gas}}
\newcommand*{\Mgas}{\mysub{M}{gas}}
\newcommand*{\Mtot}{\mysub{M}{tot}}
\newcommand*{\rhoc}{\mysub{\rho}{cr}}

\defcitealias{Allen08}{A08}
\defcitealias{Mantz08}{M08}
\defcitealias{Mantz09}{Paper~I}
\newcommand*{\cosmopaper}{\citetalias{Mantz09}}

\newcommand*{\Chandra}{{\it{Chandra}}}
\newcommand*{\ROSAT}{{\it{ROSAT}}}

\title[Cluster growth: X-ray scaling relations]{The Observed Growth of Massive Galaxy Clusters II: X-ray Scaling Relations}
\author[A. Mantz et al.]{
  A.~Mantz,$^{1,2}$\thanks{E-mail: amantz@slac.stanford.edu} S.~W.~Allen,$^{1,2}$ H.~Ebeling,$^3$ D.~Rapetti$^{1,2}$ and A.~Drlica-Wagner$^{1,2}$\\
  $^1$Kavli Institute for Particle Astrophysics and Cosmology, Stanford University, 452 Lomita Mall, Stanford, CA 94305-4085, USA\\
  $^2$SLAC National Accelerator Laboratory, 2575 Sand Hill Road, Menlo Park, CA 94025, USA\\
  $^3$Institute for Astronomy, 2680 Woodlawn Drive, Honolulu, HI 96822, USA
}
\date{Accepted 2010 April 28. Received 2010 March 14; in original form 2009 August 30}

\begin{document}
\pagerange{\pageref{firstpage}--\pageref{lastpage}} \pubyear{2010}
\maketitle
\label{firstpage}

\begin{abstract}
  This is the second in a series of papers in which we derive simultaneous constraints on cosmology and X-ray scaling relations using observations of massive, X-ray flux-selected galaxy clusters. The data set consists of 238 clusters with 0.1--2.4\keV{} luminosities $>2.5\E{44}h_{70}^{-2}\erg\second^{-1}$, and incorporates follow-up observations of 94 of those clusters using the \Chandra{} X-ray Observatory or \ROSAT{} (11 were observed with both). The clusters are drawn from three samples based on the \ROSAT{} All-Sky Survey: the \ROSAT{} Brightest Cluster Sample (78/37 clusters detected/followed-up), the \ROSAT{}-ESO Flux-Limited X-ray sample (126/25), and the bright sub-sample of the Massive Cluster Survey (34/32). Our analysis accounts self-consistently for all selection effects, covariances and systematic uncertainties. Here we describe the reduction of the follow-up X-ray observations, present results on the cluster scaling relations, and discuss their implications. Our constraints on the luminosity--mass and temperature--mass relations, measured within $r_{500}$, lead to three important results. First, the data support the conclusion that excess heating of the intracluster medium (or a combination of heating and condensation of the coldest gas) has altered its thermodynamic state from that expected in a simple, gravitationally dominated system; however, this excess heat is primarily limited to the central regions of clusters ($r<0.15r_{500}$). Second, the intrinsic scatter in the center-excised luminosity--mass relation is remarkably small, being bounded at the $<10$ per cent level in current data; for the hot, massive clusters under investigation, this scatter is smaller than in either the temperature--mass or $Y_X$--mass relations (10--15 per cent). Third, the evolution with redshift of the scaling relations is consistent with the predictions of simple, self-similar models of gravitational collapse, indicating that the mechanism responsible for heating the central regions of clusters was in operation before redshift 0.5 (the limit of our data) and that its effects on global cluster properties have not evolved strongly since then. Our results provide a new benchmark for comparison with numerical simulations of cluster formation and evolution.
\end{abstract}


\section{Introduction} \label{sec:introduction}

Establishing the relationship between total mass and observable quantities is a crucial step in deriving cosmological constraints from the growth of cosmic structure using galaxy clusters. Not only can these scaling relations provide useful proxies for mass, but they are also fundamentally important in accounting for selection effects such as Eddington bias and Malmquist bias.

The construction of X-ray flux-selected cluster samples out to redshift $z=0.5$ and beyond has now enabled investigations of dark energy using these data (\citealt{Mantz08}, hereafter \citetalias{Mantz08}; \citealt{Vikhlinin09a}). However, the task of calibrating X-ray scaling relations has become correspondingly more complex; the evolution  with cosmic time of the scaling relations and their scatter must be well understood, since such evolution can be degenerate with the effects of dark energy. The effect on cosmological constraints of systematic uncertainties in the scaling relations has been discussed in the context of future surveys by, e.g., \citet{Sahlen09}.

Fortunately, as we describe below, it is possible to simultaneously constrain both the evolution of the scaling relations and cosmological parameters, using a flux-limited sample of which some clusters have been targeted by detailed, follow-up X-ray observations. To distinguish X-ray methods from other measures of the growth of cosmic structure, including those using optically selected clusters, we refer to the resulting data set as the cluster X-ray luminosity function (XLF), although, strictly speaking, it contains a great deal more information than the luminosity function alone.

This is the second of a series of papers in which we address these issues. In a companion paper \citep[][hereafter \cosmopaper{}]{Mantz09} we describe the statistical methods required to simultaneously constrain the scaling relations and cosmology, self-consistently accounting for all selection effects\footnote{Throughout this paper, we refer to ``selection effects'' or ``selection biases'' relative to a mass-limited sample, since we are primarily interested in deriving scaling relations as a function of cluster mass.} and systematic uncertainties, and present the cosmological results from our analysis. This paper focuses on the reduction of the follow-up X-ray observations, and the constraints on the scaling relations from the simultaneous analysis. In Papers~III \citep{Rapetti09a} and IV \citep*{Mantz09b}, we respectively apply our analysis to investigations of modified gravity and neutrino properties.

In addition to their utility for cosmological investigations, cluster scaling relations are of significant astrophysical importance. Of primary interest is the heating mechanism that prevents cooling gas in dense cluster cores from condensing into stars and molecular gas at much higher rates than are observed \citep[for reviews, see][]{Peterson06,McNamara07}. The shape and evolution of the scaling relations, and specifically any departures from the simplest predictions for gravitationally dominated systems, can provide information on the physical mechanisms responsible for averting strong cooling and star formation.

The details of the cluster sample selection, the follow-up observations, and their reduction are discussed in \secref{sec:data}. \secrefs{sec:model} describes the scaling relation model and \secref{sec:analysis} our statistical method, which is more comprehensively detailed in \cosmopaper{}. In \secref{sec:results}, we present constraints on the scaling relations, and investigate various extensions to the simplest model, including possible evolution with redshift and asymmetric scatter. \secref{sec:mlce} contains a discussion of the influence of cool, X-ray bright gas in cluster centers on the scaling relations. Implications of our results for the astrophysics of the intracluster medium (ICM) are discussed in \secref{sec:astroimp}.

Unless otherwise noted, masses, luminosities and distances in tables and figures are calculated with respect to a reference cosmology, defined to be spatially flat, with Hubble constant $h=H_0/100\km\second^{-1}\Mpc^{-1}=0.7$, present mean matter density with respect to the critical density $\Omegam=0.3$, and dark energy in the form of a cosmological constant. We adopt the conventional definition of cluster radius in terms of the critical density of the Universe; thus, $r_\Delta$ is the radius within which the mean density of the cluster is $\Delta$ times the critical density at the redshift of the cluster.

\section{Data} \label{sec:data}

\subsection{Cluster samples} \label{sec:surveys}

\begin{figure}
  \centering
  \includegraphics[angle=270]{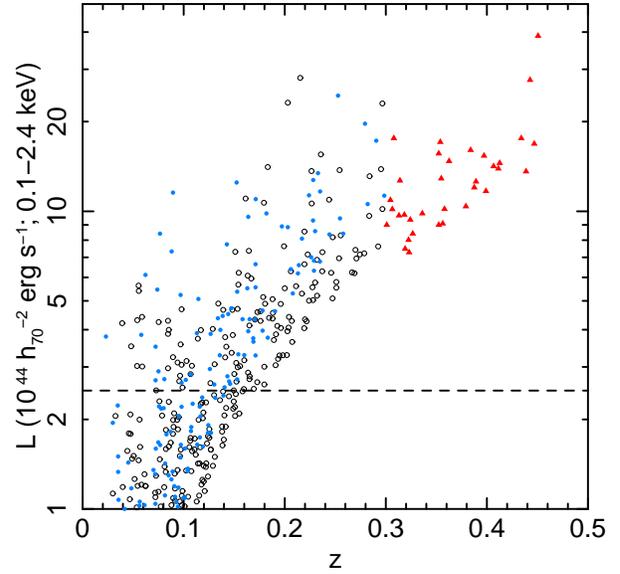}
  \caption{Luminosity--redshift distribution of clusters in the BCS (blue, filled circles), REFLEX (black, open circles) and Bright MACS (red triangles) samples which are above the respective flux limits (see text). The adopted minimum luminosity of $2.5\E{44} h_{70}^{-2}\erg\second^{-1}$ is indicated by the dashed line. Luminosities in this plot are calculated from survey fluxes, assuming our reference cosmology, and refer to the 0.1--2.4\keV{} band in the cluster rest frame. Error bars are not shown.}
  \label{fig:samples}
\end{figure}

Our data are drawn from three wide-area cluster samples derived from the \ROSAT{} All-Sky Survey \citep[RASS;][]{Trumper93}: the \ROSAT{} Brightest Cluster Sample \citep[BCS;][]{Ebeling98}, the \ROSAT{}-ESO Flux-Limited X-ray sample \citep[REFLEX;][]{Bohringer04}, and the bright sub-sample of the Massive Cluster Survey (Bright MACS; \citealt*{Ebeling01}; \citealt{Ebeling10}). Each sample covers a distinct volume of the Universe: the BCS covers the northern sky at $z<0.3$, REFLEX covers the southern sky\footnote{The coverage of REFLEX actually extends slightly into the northern hemisphere. In this study, we restrict REFLEX to declinations $<0$ and adjust the sky coverage fraction accordingly.} at $z<0.3$, and Bright MACS covers higher redshifts, $0.3<z<0.5$, at declinations $>-40\degrees$. The distributions of cluster detections from each sample in redshift and luminosity are shown in \figref{fig:samples} (here luminosities in the rest frame 0.1--2.4\keV{} band are calculated straightforwardly from the survey fluxes, assuming our reference cosmology).

The flux limits, \mysub{F}{lim}, defining the samples are respectively $4.4\E{-12}$, $3.0\E{-12}$ and $2.0\E{-12}\erg \second^{-1} \cm^{-2}$ in the 0.1--2.4~keV \ROSAT{} energy band. In order to restrict the data set to massive clusters, for which the scaling relations can be described simply, we select only clusters whose flux in this band satisfies
\begin{equation}
  \label{eq:fluxlimit}
  F \geq \mathrm{max} \left[ \mysub{F}{min}(z), ~ \mysub{F}{lim} \right],
\end{equation}
where \newpage
\begin{eqnarray}
  \label{eq:flimfcn}
  \logTen \frac{\mysub{F}{min}(z)}{\erg\second^{-1}\cm^{-2}} &=& -13.1931 - 2.31671\logTen z \nonumber \\
  & & - 0.10348\left(\logTen z\right)^2
\end{eqnarray}
approximately corresponds to a fixed, intrinsic luminosity of $2.5\E{44}h_{70}^{-2}\erg \second^{-1}$ in the 0.1--2.4\keV{} band for our reference cosmology. Using this selection criterion, the BCS, REFLEX and MACS samples respectively contribute 78, 126 and 34 clusters to our data set.\footnote{Our previous analysis in \citetalias{Mantz08} included 130 REFLEX sources using similar selection criteria. Published catalog entries RX~J0117.8-5455 and RX~J2251.7-3206 were eliminated from the present study, since high-resolution \Chandra{} imaging reveals their emission to be point-like rather than extended. Source RX~J0507.6-0238 satisfies our criteria after correcting its redshift (Dale Kocevski, private communication) from that published in \citet{Bohringer04}, and was therefore added. Three other sources (all Abell clusters), RX~J0552.8-2103, RX~J1959.1-3454 and RX~J2331.2-3630, were removed because they do not satisfy the  flux limit defined in \eqnrefs{eq:fluxlimit} and \ref{eq:flimfcn}, reflecting a small difference between our K-correction and that used by \citet{Bohringer04}, and the fact that \eqnref{eq:flimfcn} is not precisely identical to a fixed luminosity. The BCS and MACS sources selected are the same as in \citetalias{Mantz08}.} Within the redshift and flux range defined above, all three samples are thought to be approximately 100 per cent complete and pure.\footnote{In the sense that the published selection function for each sample, the likelihood for cluster detection as a function of redshift and flux, is accurate. See additional comments in \citetalias{Mantz08}.}

\subsection{Follow-up X-ray observations} \label{sec:pointed}

Additional information about a subset of the flux-selected clusters is available from follow-up X-ray observations. These observations enable more precise measurements of cluster luminosities, as well as measurements of morphology, temperature, and gas mass, which is an excellent proxy for total mass.

We incorporate \Chandra{} data for clusters at $z>0.2$ where it is available. Below this redshift, the \Chandra{} Advanced CCD Imaging Spectrometer (ACIS) field of view is typically too small to allow measurements at $r_{500}$, a canonical radius at which simulations indicate that the scaling relations are well behaved \citep{Evrard08}. For $z<0.2$, we do, however, incorporate follow-up \ROSAT{} Position Sensitive Proportional Counter (PSPC) observations. The \ROSAT{} and \Chandra{} observations were cross-calibrated using clusters in the redshift range $0.2<z<0.31$ that were observed with both instruments (\secref{sec:crosscal}). Details of the follow-up observations can be found in \tabrefs{tab:bcsobs}--\ref{tab:macsobs}; the total good exposure time of the follow-up data, including both \ROSAT{} and \Chandra{} observations, is 3.3~Ms, distributed over 94 clusters. The reduction and analysis of these data are described in the following sections.

\begin{table*}
  \centering
  \caption{Details of the follow-up observations of BCS clusters. The detector column indicates whether the observation used the \ROSAT{} PSPC or the one of the \Chandra{} ACIS CCD cameras.}
  \begin{tabular}{|lccc@{~}c@{~}clcr@{.}l|}
    \hline
    Name & RA (J2000) & Dec (J2000) & \multicolumn{3}{c}{Date} & Detector & Mode & \multicolumn{2}{c}{Exposure (ks)} \\
    \hline
Abell 2256       &  17  04  02.3  &  +78  38  14  & 1990 & Jun & 17 &  PSPC      &          &  \hspace{5ex}16 & 4   \\
Abell 1795       &  13  48  53.1  &  +26  35  38  & 1991 & Jul & 01 &  PSPC      &          &  15 & 1   \\
                 &                &               & 1992 & Jan & 04 &  PSPC      &          &  33 & 1   \\
Abell 401        &  02  58  57.2  &  +13  34  46  & 1992 & Jan & 23 &  PSPC      &          &  5 & 3    \\
                 &                &               & 1992 & Jul & 30 &  PSPC      &          &  6 & 9    \\
Abell 2029       &  15  10  55.9  &  +05  44  44  & 1992 & Aug & 10 &  PSPC      &          &  8 & 7    \\
Abell 2255       &  17  12  50.0  &  +64  03  43  & 1993 & Aug & 24 &  PSPC      &          &  11 & 8   \\
Abell 478        &  04  13  25.2  &  +10  28  00  & 1991 & Aug & 31 &  PSPC      &          &  21 & 9   \\
Abell 2142       &  15  58  21.1  &  +27  13  45  & 1992 & Aug & 25 &  PSPC      &          &  5 & 9    \\
                 &                &               & 1992 & Aug & 26 &  PSPC      &          &  4 & 8    \\
                 &                &               & 1993 & Jul & 23 &  PSPC      &          &  4 & 4    \\
Abell 2244       &  17  02  42.0  &  +34  03  24  & 1992 & Sep & 21 &  PSPC      &          &  2 & 9    \\
Abell 2034       &  15  10  12.6  &  +33  30  45  & 1993 & Jan & 31 &  PSPC      &          &  4 & 9    \\
Abell 1068       &  10  40  44.4  &  +39  57  13  & 1992 & Nov & 30 &  PSPC      &          &  9 & 8    \\
Abell 2204       &  16  32  47.2  &  +05  34  33  & 1992 & Sep & 04 &  PSPC      &          &  5 & 2    \\
Abell 2218       &  16  35  53.0  &  +66  12  36  & 1991 & May & 25 &  PSPC      &          &  31 & 9   \\
Abell 1914       &  14  26  01.0  &  +37  49  37  & 1992 & Jul & 20 &  PSPC      &          &  6 & 3    \\
Abell 665        &  08  30  58.1  &  +65  51  02  & 1991 & Apr & 10 &  PSPC      &          &  32 & 0   \\
Abell 520        &  04  54  09.0  &  +02  55  18  & 1993 & Sep & 06 &  PSPC      &          &  4 & 8    \\
                 &                &               & 2000 & Oct & 10 &  ACIS$-$I  &  VFAINT  &  9 & 4    \\
                 &                &               & 2003 & Dec & 04 &  ACIS$-$I  &  VFAINT  &  57 & 8   \\
                 &                &               & 2007 & Jan & 01 &  ACIS$-$I  &  VFAINT  &  5 & 1    \\
Abell 963        &  10  17  03.6  &  +39  02  52  & 1993 & Oct & 28 &  PSPC      &          &  9 & 1    \\
                 &                &               & 2000 & Oct & 11 &  ACIS$-$S  &  FAINT   &  36 & 3   \\
                 &                &               & 2007 & Feb & 18 &  ACIS$-$I  &  VFAINT  &  5 & 0    \\
RX J0439.0+0520  &  04  39  02.2  &  +05  20  42  & 2000 & Aug & 29 &  ACIS$-$I  &  VFAINT  &  9 & 6    \\
                 &                &               & 2007 & Nov & 12 &  ACIS$-$I  &  VFAINT  &  16 & 2   \\
                 &                &               & 2007 & Nov & 15 &  ACIS$-$I  &  VFAINT  &  7 & 9    \\
Abell 1423       &  11  57  17.4  &  +33  36  40  & 2000 & Jul & 07 &  ACIS$-$I  &  VFAINT  &  9 & 9    \\
Zwicky 2701      &  09  52  49.2  &  +51  53  06  & 2001 & Nov & 04 &  ACIS$-$S  &  VFAINT  &  25 & 9   \\
Abell 773        &  09  17  52.7  &  +51  43  36  & 1991 & Oct & 28 &  PSPC      &          &  11 & 0   \\
                 &                &               & 2000 & Sep & 05 &  ACIS$-$I  &  VFAINT  &  10 & 9   \\
                 &                &               & 2003 & Jan & 25 &  ACIS$-$I  &  VFAINT  &  7 & 6    \\
                 &                &               & 2004 & Jan & 21 &  ACIS$-$I  &  VFAINT  &  19 & 8   \\
Abell 2261       &  17  22  27.0  &  +32  07  58  & 1999 & Dec & 11 &  ACIS$-$I  &  VFAINT  &  6 & 7    \\
                 &                &               & 2004 & Jan & 14 &  ACIS$-$I  &  VFAINT  &  23 & 3   \\
Abell 1682       &  13  06  50.7  &  +46  33  30  & 2002 & Oct & 19 &  ACIS$-$I  &  VFAINT  &  1 & 8    \\
Abell 1763       &  13  35  19.0  &  +40  59  59  & 1992 & Jun & 23 &  PSPC      &          &  12 & 7   \\
                 &                &               & 2003 & Aug & 28 &  ACIS$-$I  &  VFAINT  &  19 & 3   \\
Abell 2219       &  16  40  20.3  &  +46  42  30  & 1993 & Aug & 02 &  PSPC      &          &  8 & 5    \\
                 &                &               & 2000 & Mar & 31 &  ACIS$-$S  &  FAINT   &  42 & 2   \\
Abell 2111       &  15  39  41.1  &  +34  25  07  & 1993 & Jul & 23 &  PSPC      &          &  6 & 7    \\
                 &                &               & 2000 & Mar & 22 &  ACIS$-$I  &  FAINT   &  9 & 5    \\
Zwicky 5247      &  12  34  22.1  &  +09  47  05  & 2000 & Mar & 23 &  ACIS$-$I  &  VFAINT  &  9 & 3    \\
Abell 267        &  01  52  42.2  &  +01  00  30  & 1999 & Oct & 16 &  ACIS$-$I  &  FAINT   &  7 & 1    \\
                 &                &               & 2003 & Dec & 07 &  ACIS$-$I  &  VFAINT  &  17 & 8   \\
Abell 2390       &  21  53  37.1  &  +17  41  45  & 1993 & Nov & 13 &  PSPC      &          &  8 & 4    \\
                 &                &               & 1999 & Nov & 05 &  ACIS$-$S  &  FAINT   &  8 & 0    \\
                 &                &               & 2000 & Oct & 08 &  ACIS$-$S  &  FAINT   &  9 & 8    \\
                 &                &               & 2003 & Sep & 11 &  ACIS$-$S  &  VFAINT  &  74 & 6   \\
Zwicky 2089      &  09  00  36.8  &  +20  53  40  & 2006 & Dec & 23 &  ACIS$-$I  &  VFAINT  &  9 & 1    \\
RX J2129.6+0005  &  21  29  39.7  &  +00  05  18  & 2000 & Oct & 21 &  ACIS$-$I  &  VFAINT  &  8 & 7    \\
RX J0439.0+0715  &  04  39  00.6  &  +07  16  00  & 1999 & Oct & 16 &  ACIS$-$I  &  FAINT   &  6 & 3    \\
                 &                &               & 1999 & Oct & 16 &  ACIS$-$I  &  VFAINT  &  1 & 6    \\
                 &                &               & 2003 & Jan & 04 &  ACIS$-$I  &  VFAINT  &  19 & 0   \\
Abell 1835       &  14  01  01.9  &  +02  52  40  & 1993 & Jul & 03 &  PSPC      &          &  6 & 1    \\
                 &                &               & 1994 & Jul & 05 &  PSPC      &          &  2 & 5    \\
                 &                &               & 2005 & Dec & 07 &  ACIS$-$I  &  VFAINT  &  36 & 3   \\
                 &                &               & 2006 & Jul & 24 &  ACIS$-$I  &  VFAINT  &  39 & 5   \\
                 &                &               & 2006 & Aug & 25 &  ACIS$-$I  &  VFAINT  &  117 & 9  \\
Abell 68         &  00  37  05.9  &  +09  09  36  & 2002 & Sep & 07 &  ACIS$-$I  &  VFAINT  &  10 & 0   \\
    \hline
  \end{tabular}
  \label{tab:bcsobs}
\end{table*}

\begin{table*}
  \centering
  \contcaption{}
  \begin{tabular}{|lccc@{~}c@{~}clcr@{.}l|}
    \hline
    Name & RA (J2000) & Dec (J2000) & \multicolumn{3}{c}{Date} & Detector & Mode & \multicolumn{2}{c}{Exposure (ks)} \\
    \hline
MS J1455.0+2232  &  14  57  15.1  &  +22  20  33  & 2000 & May & 19 &  ACIS$-$I  &  FAINT   &  \hspace{5ex}9 & 9    \\
                 &                &               & 2003 & Sep & 05 &  ACIS$-$I  &  VFAINT  &  91 & 9   \\
                 &                &               & 2007 & Mar & 23 &  ACIS$-$I  &  VFAINT  &  7 & 1    \\
Abell 697        &  08  42  57.6  &  +36  21  57  & 1991 & Apr & 10 &  PSPC      &          &  9 & 6    \\
                 &                &               & 2002 & Dec & 15 &  ACIS$-$I  &  VFAINT  &  17 & 5   \\
Zwicky 3146      &  10  23  39.6  &  +04  11  12  & 1993 & Nov & 13 &  PSPC      &          &  8 & 1    \\
                 &                &               & 2000 & May & 10 &  ACIS$-$I  &  FAINT   &  42 & 4   \\
                 &                &               & 2008 & Jan & 18 &  ACIS$-$I  &  VFAINT  &  31 & 2   \\
Abell 781        &  09  20  25.3  &  +30  30  11  & 2000 & Oct & 03 &  ACIS$-$I  &  VFAINT  &  8 & 7    \\
    \hline
  \end{tabular}
\end{table*}

\begin{table*}
  \centering
  \caption{Details of the follow-up observations of REFLEX clusters (see caption for \tabref{tab:bcsobs}).}
  \begin{tabular}{|lccc@{~}c@{~}clcr@{.}l|}
    \hline
    Name & RA (J2000) & Dec (J2000) & \multicolumn{3}{c}{Date} & Detector & Mode & \multicolumn{2}{c}{Exposure (ks)} \\
    \hline
Abell 3558       &  13  27  58.4  &  $-$31  30  04  & 1991 & Jul & 17 &  PSPC      &          &  \hspace{5ex}24 & 0  \\
Abell 85         &  00  41  50.1  &  $-$09  18  22  & 1992 & Jun & 11 &  PSPC      &          &  3 & 2   \\
                 &                &               & 1992 & Jul & 01 &  PSPC      &          &  8 & 3   \\
Abell 3667       &  20  12  31.2  &  $-$56  49  47  & 1992 & Oct & 09 &  PSPC      &          &  6 & 3   \\
Abell 3266       &  04  31  21.5  &  $-$61  26  21  & 1992 & Apr & 30 &  PSPC      &          &  6 & 6   \\
                 &                &               & 1993 & Aug & 19 &  PSPC      &          &  13 & 2  \\
Abell 3112       &  03  17  58.2  &  $-$44  14  07  & 1992 & Dec & 17 &  PSPC      &          &  7 & 0   \\
Abell 2597       &  23  25  19.6  &  $-$12  07  27  & 1991 & Nov & 27 &  PSPC      &          &  6 & 7   \\
Abell 3921       &  22  49  57.0  &  $-$64  25  53  & 1992 & Nov & 15 &  PSPC      &          &  11 & 5  \\
MS J1111.8       &  11  14  12.7  &  $-$38  11  25  & 1993 & Jan & 11 &  PSPC      &          &  18 & 5  \\
Abell 1689       &  13  11  29.6  &  $-$01  20  28  & 1992 & Jul & 18 &  PSPC      &          &  13 & 5  \\
Abell 2163       &  16  15  46.0  &  $-$06  08  54  & 1992 & Feb & 28 &  PSPC      &          &  3 & 1   \\
                 &                &               & 1992 & Sep & 01 &  PSPC      &          &  7 & 0   \\
                 &                &               & 2000 & Jul & 29 &  ACIS$-$I  &  VFAINT  &  9 & 4   \\
                 &                &               & 2001 & Jun & 16 &  ACIS$-$I  &  VFAINT  &  67 & 4  \\
Abell 209        &  01  31  53.1  &  $-$13  36  48  & 2000 & Sep & 09 &  ACIS$-$I  &  VFAINT  &  8 & 7   \\
                 &                &               & 2003 & Aug & 03 &  ACIS$-$I  &  VFAINT  &  9 & 6   \\
RX J1504.1$-$0248  &  15  04  07.6  &  $-$02  48  16  & 2004 & Jan & 07 &  ACIS$-$I  &  FAINT   &  12 & 0  \\
                 &                &               & 2005 & Mar & 20 &  ACIS$-$I  &  VFAINT  &  33 & 8  \\
RX J0304.1$-$3656  &  03  04  03.3  &  $-$36  56  30  & 2008 & Mar & 16 &  ACIS$-$I  &  VFAINT  &  19 & 9  \\
RX J0237.4$-$2630  &  02  37  27.4  &  $-$26  30  28  & 2008 & Mar & 03 &  ACIS$-$I  &  VFAINT  &  7 & 4   \\
Abell 2667       &  23  51  39.7  &  $-$26  04  60  & 1992 & Dec & 05 &  PSPC      &          &  2 & 0   \\
                 &                &               & 1992 & Dec & 05 &  PSPC      &          &  3 & 0   \\
                 &                &               & 2001 & Jun & 19 &  ACIS$-$S  &  VFAINT  &  9 & 4   \\
RX J0638.7$-$5358  &  06  38  47.3  &  $-$53  58  28  & 2008 & Apr & 11 &  ACIS$-$I  &  VFAINT  &  19 & 9  \\
RX J0220.9$-$3829  &  02  20  56.5  &  $-$38  28  52  & 2008 & Feb & 29 &  ACIS$-$I  &  VFAINT  &  19 & 9  \\
Abell 521        &  04  54  07.4  &  $-$10  13  24  & 1999 & Dec & 23 &  ACIS$-$I  &  VFAINT  &  36 & 8  \\
                 &                &               & 2000 & Oct & 13 &  ACIS$-$S  &  VFAINT  &  18 & 6  \\
RX J0307.0$-$2840  &  03  07  02.0  &  $-$28  39  56  & 2008 & Mar & 13 &  ACIS$-$I  &  VFAINT  &  18 & 1  \\
RX J2011.3$-$5725  &  20  11  27.2  &  $-$57  25  10  & 2004 & Jun & 08 &  ACIS$-$I  &  VFAINT  &  22 & 3  \\
RX J0232.2$-$4420  &  02  32  17.7  &  $-$44  20  55  & 2004 & Jun & 08 &  ACIS$-$I  &  VFAINT  &  7 & 9   \\
RX J0528.9$-$3927  &  05  28  53.3  &  $-$39  28  19  & 2004 & Mar & 10 &  ACIS$-$I  &  VFAINT  &  15 & 9  \\
RX J0043.4$-$2037  &  00  43  24.8  &  $-$20  37  24  & 2008 & Feb & 02 &  ACIS$-$I  &  VFAINT  &  19 & 7  \\
1ES 0657$-$558     &  06  58  27.5  &  $-$55  56  32  & 1997 & Feb & 28 &  PSPC      &          &  4 & 5   \\
                 &                &               & 2000 & Oct & 16 &  ACIS$-$I  &  FAINT   &  25 & 3  \\
                 &                &               & 2002 & Jul & 12 &  ACIS$-$I  &  VFAINT  &  81 & 1  \\
                 &                &               & 2004 & Aug & 10 &  ACIS$-$I  &  VFAINT  &  21 & 3  \\
                 &                &               & 2004 & Aug & 11 &  ACIS$-$I  &  VFAINT  &  93 & 9  \\
                 &                &               & 2004 & Aug & 14 &  ACIS$-$I  &  VFAINT  &  77 & 5  \\
                 &                &               & 2004 & Aug & 15 &  ACIS$-$I  &  VFAINT  &  31 & 2  \\
                 &                &               & 2004 & Aug & 17 &  ACIS$-$I  &  VFAINT  &  79 & 8  \\
                 &                &               & 2004 & Aug & 19 &  ACIS$-$I  &  VFAINT  &  70 & 7  \\
                 &                &               & 2004 & Aug & 23 &  ACIS$-$I  &  VFAINT  &  22 & 9  \\
                 &                &               & 2004 & Aug & 25 &  ACIS$-$I  &  VFAINT  &  39 & 1  \\
Abell 2537       &  23  08  22.0  &  $-$02  11  30  & 2004 & Sep & 09 &  ACIS$-$S  &  VFAINT  &  36 & 2  \\
    \hline
  \end{tabular}
  \label{tab:reflexobs}
\end{table*}

\begin{table*}
  \centering
  \caption{Details of the follow-up observations of Bright MACS clusters (see caption for \tabref{tab:bcsobs}). The last two observations listed are not used in this paper: the MACS~J0358 observation took place after this paper was initially submitted, and we conservatively chose to exclude the MACS~J2311 observation because it coincides with a long-duration background flare. Their derived properties are nevertheless included in \tabref{tab:mltmacs} so that the complete sample of \citet{Ebeling10} is represented.}
  \begin{tabular}{|lccc@{~}c@{~}clcr@{.}l|}
    \hline
    Name & RA (J2000) & Dec (J2000) & \multicolumn{3}{c}{Date} & Detector & Mode & \multicolumn{2}{c}{Exposure (ks)} \\
    \hline
MACS J2245.0+2637    &  22  45  04.6  &  +26    38  04  &  2002  &  Nov  &  24  &  ACIS$-$I  &  VFAINT  &  \hspace{5ex}13  &  9  \\
MACS J1131.8$-$1955  &  11  31  55.6  &  $-$19  55  45  &  1993  &  Jun  &  27  &  PSPC      &          &  7               &  2  \\
                     &                &                 &  2002  &  Jun  &  14  &  ACIS$-$I  &  VFAINT  &  13              &  1  \\
MACS J0014.3$-$3022  &  00  14  18.8  &  $-$30  23  18  &  1992  &  Jun  &  16  &  PSPC      &          &  13              &  6  \\
                     &                &                 &  2001  &  Sep  &  03  &  ACIS$-$S  &  VFAINT  &  23              &  8  \\
                     &                &                 &  2006  &  Nov  &  08  &  ACIS$-$I  &  VFAINT  &  12              &  3  \\
                     &                &                 &  2007  &  Jun  &  10  &  ACIS$-$I  &  VFAINT  &  37              &  7  \\
                     &                &                 &  2007  &  Jun  &  14  &  ACIS$-$I  &  VFAINT  &  23              &  9  \\
MACS J2140.2$-$2339  &  21  40  15.2  &  $-$23  39  40  &  1993  &  Nov  &  07  &  PSPC      &          &  9               &  2  \\
                     &                &                 &  1999  &  Nov  &  18  &  ACIS$-$S  &  VFAINT  &  40              &  5  \\
                     &                &                 &  2003  &  Nov  &  18  &  ACIS$-$S  &  VFAINT  &  24              &  0  \\
MACS J0242.5$-$2132  &  02  42  35.9  &  $-$21  32  26  &  2002  &  Feb  &  07  &  ACIS$-$I  &  VFAINT  &  9               &  6  \\
MACS J1427.6$-$2521  &  14  27  39.5  &  $-$25  21  03  &  2002  &  Jun  &  29  &  ACIS$-$I  &  VFAINT  &  14              &  6  \\
                     &                &                 &  2008  &  Jun  &  11  &  ACIS$-$I  &  VFAINT  &  26              &  3  \\
MACS J0547.0$-$3904  &  05  47  01.5  &  $-$39  04  26  &  2002  &  Oct  &  20  &  ACIS$-$I  &  VFAINT  &  20              &  9  \\
MACS J0257.6$-$2209  &  02  57  41.3  &  $-$22  09  13  &  2001  &  Nov  &  12  &  ACIS$-$I  &  VFAINT  &  20              &  5  \\
MACS J2049.9$-$3217  &  20  49  55.3  &  $-$32  16  49  &  2002  &  Dec  &  08  &  ACIS$-$I  &  VFAINT  &  22              &  9  \\
MACS J2229.7$-$2755  &  22  29  45.2  &  $-$27  55  36  &  2002  &  Nov  &  13  &  ACIS$-$I  &  VFAINT  &  12              &  9  \\
                     &                &                 &  2007  &  Dec  &  09  &  ACIS$-$I  &  VFAINT  &  13              &  3  \\
MACS J1319.9+7003    &  13  20  07.5  &  +70    04  37  &  2002  &  Sep  &  15  &  ACIS$-$I  &  VFAINT  &  8               &  7  \\
MACS J0520.7$-$1328  &  05  20  42.2  &  $-$13  28  47  &  2002  &  Feb  &  10  &  ACIS$-$I  &  VFAINT  &  19              &  0  \\
MACS J1931.8$-$2634  &  19  31  49.6  &  $-$26  34  34  &  2002  &  Oct  &  20  &  ACIS$-$I  &  VFAINT  &  13              &  1  \\
MACS J0035.4$-$2015  &  00  35  26.2  &  $-$20  15  46  &  2003  &  Jan  &  22  &  ACIS$-$I  &  VFAINT  &  20              &  8  \\
MACS J0947.2+7623    &  09  47  13.0  &  +76    23  14  &  2000  &  Oct  &  20  &  ACIS$-$I  &  VFAINT  &  11              &  7  \\
MACS J1115.8+0129    &  11  15  51.9  &  +01    29  55  &  2003  &  Jan  &  23  &  ACIS$-$I  &  VFAINT  &  8               &  7  \\
                     &                &                 &  2008  &  Feb  &  03  &  ACIS$-$I  &  VFAINT  &  34              &  7  \\
MACS J0308.9+2645    &  03  08  56.0  &  +26    45  35  &  2002  &  Mar  &  10  &  ACIS$-$I  &  VFAINT  &  23              &  4  \\
MACS J0404.6+1109    &  04  04  32.7  &  +11    08  11  &  2002  &  Feb  &  20  &  ACIS$-$I  &  VFAINT  &  17              &  6  \\
MACS J1532.8+3021    &  15  32  53.8  &  +30    20  59  &  2001  &  Aug  &  26  &  ACIS$-$S  &  VFAINT  &  9               &  4  \\
                     &                &                 &  2001  &  Sep  &  06  &  ACIS$-$I  &  VFAINT  &  10              &  0  \\
MACS J0011.7$-$1523  &  00  11  42.8  &  $-$15  23  22  &  2002  &  Nov  &  20  &  ACIS$-$I  &  VFAINT  &  20              &  8  \\
                     &                &                 &  2005  &  Jun  &  28  &  ACIS$-$I  &  VFAINT  &  37              &  2  \\
MACS J0949.8+1708    &  09  49  51.8  &  +17    07  08  &  2002  &  Nov  &  06  &  ACIS$-$I  &  VFAINT  &  14              &  3  \\
MACS J1720.2+3536    &  17  20  16.7  &  +35    36  23  &  2002  &  Nov  &  03  &  ACIS$-$I  &  VFAINT  &  19              &  8  \\
                     &                &                 &  2005  &  Nov  &  22  &  ACIS$-$I  &  VFAINT  &  29              &  5  \\
MACS J1731.6+2252    &  17  31  39.2  &  +22    51  50  &  2002  &  Nov  &  03  &  ACIS$-$I  &  VFAINT  &  18              &  5  \\
MACS J2211.7$-$0349  &  22  11  45.9  &  $-$03  49  42  &  2002  &  Oct  &  08  &  ACIS$-$I  &  VFAINT  &  15              &  2  \\
MACS J0429.6$-$0253  &  04  29  36.0  &  $-$02  53  06  &  2002  &  Feb  &  07  &  ACIS$-$I  &  VFAINT  &  22              &  1  \\
MACS J0159.8$-$0849  &  01  59  49.4  &  $-$08  49  60  &  2002  &  Oct  &  02  &  ACIS$-$I  &  VFAINT  &  17              &  4  \\
                     &                &                 &  2004  &  Dec  &  04  &  ACIS$-$I  &  VFAINT  &  35              &  3  \\
MACS J2228.5+2036    &  22  28  32.8  &  +20    37  15  &  2003  &  Jan  &  22  &  ACIS$-$I  &  VFAINT  &  19              &  9  \\
MACS J0152.5$-$2852  &  01  52  33.9  &  $-$28  53  33  &  2002  &  Sep  &  17  &  ACIS$-$I  &  VFAINT  &  17              &  3  \\
MACS J1206.2$-$0847  &  12  06  12.3  &  $-$08  48  06  &  2002  &  Dec  &  15  &  ACIS$-$I  &  VFAINT  &  23              &  5  \\
MACS J0417.5$-$1154  &  04  17  34.3  &  $-$11  54  27  &  2002  &  Mar  &  10  &  ACIS$-$I  &  VFAINT  &  11              &  2  \\
MACS J2243.3$-$0935  &  22  43  21.4  &  $-$09  35  43  &  2002  &  Dec  &  23  &  ACIS$-$I  &  VFAINT  &  20              &  2  \\
MACS J1347.5$-$1144  &  13  47  30.8  &  $-$11  45  09  &  2000  &  Mar  &  05  &  ACIS$-$S  &  VFAINT  &  8               &  9  \\
                     &                &                 &  2000  &  Apr  &  29  &  ACIS$-$S  &  FAINT   &  10              &  0  \\
                     &                &                 &  2001  &  May  &  10  &  ACIS$-$S  &  VFAINT  &  89              &  6  \\
                     &                &                 &  2003  &  Sep  &  03  &  ACIS$-$I  &  VFAINT  &  55  &  8 \vspace{2mm} \\
MACS J0358.8$-$2955  &  03  58  51.2  &  $-$29  55  22  &  2009  &  Oct  &  18  &  ACIS$-$I  &  VFAINT  &  8               &  7  \\
MACS J2311.5+0338    &  23  11  35.3  &  +03    38  25  &  2002  &  Sep  &  07  &  ACIS$-$I  &  VFAINT  &  6               &  5  \\
    \hline
  \end{tabular}
  \label{tab:macsobs}
\end{table*}

\subsubsection{\Chandra{} data analysis} \label{sec:chandraobs}
The standard level-1 event files distributed by the \Chandra{} X-ray Center (CXC) were reprocessed in accordance with CXC recommendations, using the {\sc ciao} software package\footnote{{\url{http://cxc.harvard.edu/ciao/}}} (version 4.1.1, CALDB 4.1.2). This processing includes removal of bad pixels, corrections for cosmic ray afterglows and charge transfer inefficiency, and application of standard grade and status filters, using appropriate time-dependent gain and calibration products. The extra information available for observations taken in VFAINT mode (the majority) was used to improve cosmic ray rejection. The data were cleaned to remove times of high or unstable background using the energy ranges and time bins recommended by the CXC. Blank-field data sets made available by the CXC were tailored to each observation and cleaned in an identical manner to the real data. The normalizations of these blank-sky files were scaled to match the count rates in the target observations measured in the 9.5--12~keV band.

After identifying and masking point sources, flat-fielded images and background-subtracted surface brightness profiles were prepared in the 0.7--2.0\keV{} energy band. The emissivity in this band is largely insensitive to the gas temperature, $kT$, provided $kT>3\keV$ (as is the case for all the clusters studied here), making it the preferred energy range for determining the gas mass. The center of each cluster was identified with the centroid of 0.7--2.0\keV{} emission after masking point sources.

Our spectral analysis is a two-stage process. In the first stage, spectra in the 0.8--7.0\keV{} band were extracted in an annulus about the cluster centers. The inner radii of all annuli were set to 100\kpc{} in order to prevent relatively cool gas in the cores from strongly influencing the spectral fits. The outer radii were initially chosen to be the radii at which the signal-to-noise ratio apparent in the 0.8--7.0\keV{} surface brightness profiles falls to 2. Background spectra were extracted from the blank-sky fields for targets at $z<0.3$, and from source-free regions of the detector for $z>0.3$ targets. When required, e.g. due to the presence of strong excess soft emission in the field, a model for additional, soft thermal emission was included in the spectral modeling of the background. Photon-weighted response matrices and effective area files were generated for each observation using calibration files for the appropriate period.

The spectral analysis was performed using {\sc xspec}.\footnote{{\url{http://heasarc.gsfc.nasa.gov/docs/xanadu/xspec/}}} The spectra were fit to a single-temperature, optically thin thermal emission model evaluated with the {\sc mekal} code (\citealt{Kaastra93}; \citealt*{Mewe95}, incorporating the Fe-L calculations of \citealt*{Liedhal95}), including a model for photoelectric absorption due to Galactic hydrogen \citep{Balucinska92}. Column densities were fixed to the Galactic values determined from the H$_\mathrm{I}$ survey of \citet{Kalberla05}, unless the published value exceeded $10^{21}\cm^{-2}$, in which case they were fit as free parameters.\footnote{We note that the choice of column density does not have a significant impact on our derived luminosities, temperatures and masses. In particular, the energy bands used ($E>0.7\keV$), and the use of locally determined backgrounds for the surface brightness analysis (\secref{sec:deproj}) and, when possible, for the spectral analysis, make our results relatively insensitive to uncertainties in the soft X-ray background. Residual uncertainties are within the systematic tolerances defined in \secref{sec:deproj}.} The abundances of all metals were assumed to vary with a common ratio, $Z$, relative to solar values \citep{Anders89}, which was a free parameter in the fit for each cluster. We minimized the modified C-statistic \citep{Cash79,Arnaud96} rather than $\chi^2$, as is appropriate for spectra with few counts per bin. Once a best fitting model was identified, the parameter space was explored using Markov Chain Monte Carlo; these Monte Carlo samples were used to propagate the uncertainty due to the fit in the conversion of 0.7--2.0\keV{} count rate to 0.1--2.4\keV{} flux for each observation. We note that, since both the count rate and flux in this conversion are at soft energies, the conversion is only weakly dependent on temperature and has negligible dependence on metallicity, provided that the temperature is sufficiently high ($kT>3\keV$).

The second stage of the spectral analysis uses an estimate for $r_{500}$, obtained as described in \secref{sec:deproj} below. New spectra in the annuli (0.15--0.5)$r_{500}$ and (0.15--1)$r_{500}$ were extracted and fit as above. For observations of nearby clusters where a large fraction of the (0.15--1)$r_{500}$ region falls outside the detector, only the (0.15--0.5)$r_{500}$ region was analyzed; we then estimated the (0.15--1)$r_{500}$ temperature using the relation $T_{0.15-1}/T_{0.15-0.5}=0.957\pm0.009$, fit from the other clusters. (This result is consistent with that of \citealt{Vikhlinin09}, who followed a similar procedure; see Section~4.1.1 and Figure~6 of that work.) This final step is motivated by the desire to measure a temperature within $r_{500}$, for consistency with the mass and luminosity measurements (\secref{sec:deproj}), while still reliably excising the cool core, if one is present. We note, however, that this level of detail is not entirely necessary; the $\sim 4$ per cent reduction in temperature within $r_{500}$ compared with the temperature measurement from the first stage of the analysis is within statistical errors for most systems and has negligible impact on the determination of masses and luminosities. We therefore do not further iterate the determination of $r_{500}$ using this new temperature estimate. The resulting temperature measurements are referred to as $\mysub{kT}{ce}$ in the following, the subscript indicating ``center-excised'', and are listed in \tabrefs{tab:mltbcs}--\ref{tab:mltmacs}.

We note that the uncertainties on temperatures measured within $r_{500}$ are necessarily somewhat larger than those on our initial estimates from higher signal-to-noise regions. Ultimately, however, this has negligible effect on our results, since this additional uncertainty is smaller than the systematic allowances associated with instrument calibration and mass estimations (\secref{sec:deproj}).

\begin{table*}
  \centering
  \caption{Redshifts and derived properties of BCS clusters from follow-up observations. Temperatures in this work are measured in the aperture (0.15--1)$r_{500}$. Luminosities are in the 0.1--2.4\keV{} band in the cluster rest frame and have been cross-calibrated to the standard of \ROSAT{} (\secref{sec:crosscal}). Gas masses measured with \Chandra{} have similarly been calibrated to the \ROSAT{} standard. Total masses are independent of this cross-calibration, but do depend on the value of \fgas{}, which is assumed here to be 0.1104 at $r_{2500}$. As discussed in the text, recent \Chandra{} calibration updates favor an increase in the measured values of \fgas{} at $r_{2500}$ by 5--10~per cent, implying reductions of 3--5~per cent in $r_{500}$ and 7--15~per cent in $\Mgas{}_{,500}$ and $M_{500}$, with respect to the values listed here. This dependence is accounted for fully in our analysis (see also \cosmopaper{}). Error bars in this table do not include contributions due to systematic uncertainty in \fgas{}, or uncertainty in the overall instrument calibration, whose effects are correlated across all clusters. $kT$ references: [1]~\citet{Horner01}, [2]~\citet{Vikhlinin09}, [3]~this work.}
  \begin{tabular}{|llr@{$\pm$}lr@{$\pm$}lr@{$\pm$}lr@{$\pm$}lr@{$~\pm~$}lr@{$\pm$}lc|}
    \hline
    Name & \multicolumn{1}{c}{$z$} & \multicolumn{2}{c}{$r_{500}$} & \multicolumn{2}{c}{$\Mgas{}_{,500}$} & \multicolumn{2}{c}{$M_{500}$} & \multicolumn{2}{c}{$L_{500}$} & \multicolumn{2}{c}{$\mysub{L}{500,ce}$} & \multicolumn{2}{c}{$\mysub{kT}{ce}$} & $kT$ \\
    & & \multicolumn{2}{c}{(Mpc)} & \multicolumn{2}{c}{$(10^{14}\Msun)$} & \multicolumn{2}{c}{$(10^{14}\Msun)$} & \multicolumn{2}{c}{$(10^{44}\erg\second^{-1})$} & \multicolumn{2}{c}{$(10^{44}\erg\second^{-1})$} & \multicolumn{2}{c}{(keV)} &  ref \\
    \hline
Abell 2256       &  0.0581  &  $1.33  ~$&$~  0.08$  &  $0.82  ~$&$~  0.09$  &  $7.2   ~$&$~  1.0$  &  $\hspace{2.5ex}4.6   ~$&$~  0.4$  &  $ 3.46$&$0.32$  &  $6.90   ~$&$~  0.16$  &  1  \\
Abell 1795       &  0.0622  &  $1.22  ~$&$~  0.06$  &  $0.63  ~$&$~  0.07$  &  $5.5   ~$&$~  0.8$  &  $6.0   ~$&$~  0.4$  &  $ 2.48$&$0.16$  &  $6.14   ~$&$~  0.10$  &  2  \\
Abell 401        &  0.0743  &  $1.49  ~$&$~  0.08$  &  $1.16  ~$&$~  0.15$  &  $10.1  ~$&$~  1.6$  &  $6.8   ~$&$~  0.4$  &  $ 4.45$&$0.29$  &  $7.72   ~$&$~  0.30$  &  2  \\
Abell 2029       &  0.0779  &  $1.45  ~$&$~  0.07$  &  $1.07  ~$&$~  0.13$  &  $9.3   ~$&$~  1.4$  &  $10.6  ~$&$~  1.0$  &  $ 4.14$&$0.39$  &  $8.22   ~$&$~  0.16$  &  2  \\
Abell 2255       &  0.0809  &  $1.24  ~$&$~  0.08$  &  $0.67  ~$&$~  0.10$  &  $5.9   ~$&$~  1.0$  &  $2.9   ~$&$~  0.3$  &  $ 2.39$&$0.22$  &  $6.42   ~$&$~  0.16$  &  1  \\
Abell 478        &  0.0881  &  $1.48  ~$&$~  0.08$  &  $1.15  ~$&$~  0.15$  &  $10.1  ~$&$~  1.6$  &  $13.3  ~$&$~  1.2$  &  $ 4.88$&$0.47$  &  $7.96   ~$&$~  0.27$  &  2  \\
Abell 2142       &  0.0904  &  $1.64  ~$&$~  0.08$  &  $1.59  ~$&$~  0.19$  &  $13.9  ~$&$~  2.1$  &  $12.4  ~$&$~  0.6$  &  $ 6.43$&$0.35$  &  $10.04  ~$&$~  0.26$  &  2  \\
Abell 2244       &  0.0989  &  $1.25  ~$&$~  0.07$  &  $0.70  ~$&$~  0.11$  &  $6.2   ~$&$~  1.1$  &  $5.2   ~$&$~  0.5$  &  $ 2.72$&$0.27$  &  $5.37   ~$&$~  0.12$  &  2  \\
Abell 2034       &  0.113   &  $1.28  ~$&$~  0.07$  &  $0.77  ~$&$~  0.10$  &  $6.7   ~$&$~  1.0$  &  $4.0   ~$&$~  0.4$  &  $ 2.88$&$0.27$  &  $7.15   ~$&$~  0.32$  &  1  \\
Abell 1068       &  0.1386  &  $1.04  ~$&$~  0.05$  &  $0.43  ~$&$~  0.06$  &  $3.7   ~$&$~  0.6$  &  $5.6   ~$&$~  0.5$  &  $ 1.78$&$0.18$  &  $3.87   ~$&$~  0.12$  &  1  \\
Abell 2204       &  0.1511  &  $1.46  ~$&$~  0.07$  &  $1.18  ~$&$~  0.14$  &  $10.3  ~$&$~  1.5$  &  $17.9  ~$&$~  1.6$  &  $ 5.30$&$0.52$  &  $8.55   ~$&$~  0.58$  &  2  \\
Abell 2218       &  0.171   &  $1.28  ~$&$~  0.09$  &  $0.82  ~$&$~  0.12$  &  $7.2   ~$&$~  1.2$  &  $5.1   ~$&$~  0.5$  &  $ 3.35$&$0.32$  &  $6.97   ~$&$~  0.37$  &  1  \\
Abell 1914       &  0.1712  &  $1.46  ~$&$~  0.10$  &  $1.21  ~$&$~  0.14$  &  $10.6  ~$&$~  1.5$  &  $11.3  ~$&$~  1.0$  &  $ 5.44$&$0.55$  &  $9.48   ~$&$~  0.49$  &  1  \\
Abell 665        &  0.1818  &  $1.55  ~$&$~  0.07$  &  $1.46  ~$&$~  0.16$  &  $12.7  ~$&$~  1.8$  &  $8.6   ~$&$~  0.8$  &  $ 5.65$&$0.52$  &  $8.03   ~$&$~  0.24$  &  1  \\
Abell 520        &  0.203   &  $1.50  ~$&$~  0.08$  &  $1.37  ~$&$~  0.15$  &  $11.9  ~$&$~  1.6$  &  $8.4   ~$&$~  0.3$  &  $ 6.41$&$0.20$  &  $7.23   ~$&$~  0.23$  &  3  \\
Abell 963        &  0.206   &  $1.25  ~$&$~  0.06$  &  $0.78  ~$&$~  0.10$  &  $6.8   ~$&$~  1.0$  &  $6.5   ~$&$~  0.2$  &  $ 3.47$&$0.15$  &  $6.08   ~$&$~  0.30$  &  3  \\
RX J0439.0+0520  &  0.208   &  $0.92  ~$&$~  0.05$  &  $0.31  ~$&$~  0.04$  &  $2.7   ~$&$~  0.5$  &  $4.4   ~$&$~  0.1$  &  $ 1.56$&$0.06$  &  $4.96   ~$&$~  0.54$  &  3  \\
Abell 1423       &  0.213   &  $1.35  ~$&$~  0.10$  &  $1.00  ~$&$~  0.20$  &  $8.7   ~$&$~  2.0$  &  $6.2   ~$&$~  0.4$  &  $ 3.97$&$0.24$  &  $5.75   ~$&$~  0.59$  &  3  \\
Zwicky 2701      &  0.214   &  $1.04  ~$&$~  0.06$  &  $0.46  ~$&$~  0.07$  &  $4.0   ~$&$~  0.7$  &  $4.5   ~$&$~  0.2$  &  $ 1.79$&$0.11$  &  $6.75   ~$&$~  0.54$  &  3  \\
Abell 773        &  0.217   &  $1.34  ~$&$~  0.06$  &  $0.98  ~$&$~  0.10$  &  $8.6   ~$&$~  1.1$  &  $7.5   ~$&$~  0.2$  &  $ 4.77$&$0.15$  &  $7.37   ~$&$~  0.45$  &  3  \\
Abell 2261       &  0.224   &  $1.59  ~$&$~  0.09$  &  $1.65  ~$&$~  0.25$  &  $14.4  ~$&$~  2.6$  &  $12.0  ~$&$~  0.4$  &  $ 5.63$&$0.23$  &  $6.10   ~$&$~  0.32$  &  3  \\
Abell 1682       &  0.226   &  $1.50  ~$&$~  0.13$  &  $1.41  ~$&$~  0.34$  &  $12.4  ~$&$~  3.2$  &  $6.5   ~$&$~  0.7$  &  $ 4.96$&$0.61$  &  $7.01   ~$&$~  2.14$  &  3  \\
Abell 1763       &  0.2279  &  $1.67  ~$&$~  0.11$  &  $1.94  ~$&$~  0.33$  &  $17.0  ~$&$~  3.4$  &  $10.5  ~$&$~  0.6$  &  $ 7.29$&$0.39$  &  $6.32   ~$&$~  0.40$  &  3  \\
Abell 2219       &  0.2281  &  $1.74  ~$&$~  0.08$  &  $2.16  ~$&$~  0.23$  &  $18.9  ~$&$~  2.5$  &  $15.5  ~$&$~  0.8$  &  $ 9.58$&$0.53$  &  $10.90  ~$&$~  0.53$  &  3  \\
Zwicky 5247      &  0.229   &  $1.31  ~$&$~  0.14$  &  $0.94  ~$&$~  0.18$  &  $8.2   ~$&$~  1.8$  &  $4.3   ~$&$~  0.3$  &  $ 3.74$&$0.24$  &  $5.31   ~$&$~  1.07$  &  3  \\
Abell 2111       &  0.229   &  $1.28  ~$&$~  0.11$  &  $0.88  ~$&$~  0.20$  &  $7.8   ~$&$~  1.9$  &  $5.0   ~$&$~  0.3$  &  $ 3.61$&$0.22$  &  $6.51   ~$&$~  0.72$  &  3  \\
Abell 267        &  0.230   &  $1.22  ~$&$~  0.07$  &  $0.76  ~$&$~  0.10$  &  $6.6   ~$&$~  1.1$  &  $5.8   ~$&$~  0.2$  &  $ 3.36$&$0.16$  &  $7.13   ~$&$~  0.71$  &  3  \\
Abell 2390       &  0.2329  &  $1.61  ~$&$~  0.07$  &  $1.73  ~$&$~  0.17$  &  $15.2  ~$&$~  1.9$  &  $17.3  ~$&$~  0.5$  &  $ 8.69$&$0.29$  &  $10.28  ~$&$~  0.38$  &  3  \\
Zwicky 2089      &  0.2347  &  $0.95  ~$&$~  0.04$  &  $0.35  ~$&$~  0.04$  &  $3.1   ~$&$~  0.4$  &  $5.8   ~$&$~  0.3$  &  $ 1.70$&$0.13$  &  $6.55   ~$&$~  1.47$  &  3  \\
RX J2129.6+0005  &  0.235   &  $1.28  ~$&$~  0.07$  &  $0.88  ~$&$~  0.12$  &  $7.7   ~$&$~  1.2$  &  $9.9   ~$&$~  0.5$  &  $ 4.17$&$0.27$  &  $6.34   ~$&$~  0.62$  &  3  \\
RX J0439.0+0715  &  0.2443  &  $1.27  ~$&$~  0.06$  &  $0.85  ~$&$~  0.10$  &  $7.4   ~$&$~  1.0$  &  $8.1   ~$&$~  0.3$  &  $ 4.22$&$0.16$  &  $6.59   ~$&$~  0.45$  &  3  \\
Abell 1835       &  0.2528  &  $1.49  ~$&$~  0.06$  &  $1.41  ~$&$~  0.12$  &  $12.3  ~$&$~  1.4$  &  $21.1  ~$&$~  0.6$  &  $ 6.73$&$0.23$  &  $9.00   ~$&$~  0.25$  &  3  \\
Abell 68         &  0.2546  &  $1.27  ~$&$~  0.07$  &  $0.87  ~$&$~  0.12$  &  $7.6   ~$&$~  1.2$  &  $6.9   ~$&$~  0.4$  &  $ 4.42$&$0.27$  &  $7.56   ~$&$~  0.97$  &  3  \\
MS J1455.0+2232  &  0.2578  &  $1.19  ~$&$~  0.06$  &  $0.71  ~$&$~  0.09$  &  $6.2   ~$&$~  1.0$  &  $11.0  ~$&$~  0.3$  &  $ 3.13$&$0.12$  &  $4.54   ~$&$~  0.16$  &  3  \\
Abell 697        &  0.282   &  $1.65  ~$&$~  0.09$  &  $1.96  ~$&$~  0.27$  &  $17.1  ~$&$~  2.9$  &  $14.4  ~$&$~  0.8$  &  $ 8.95$&$0.50$  &  $10.93  ~$&$~  1.11$  &  3  \\
Zwicky 3146      &  0.2906  &  $1.35  ~$&$~  0.06$  &  $1.08  ~$&$~  0.10$  &  $9.4   ~$&$~  1.2$  &  $19.1  ~$&$~  0.7$  &  $ 5.82$&$0.26$  &  $8.38   ~$&$~  0.44$  &  3  \\
Abell 781        &  0.2984  &  $1.26  ~$&$~  0.07$  &  $0.90  ~$&$~  0.14$  &  $7.9   ~$&$~  1.4$  &  $6.0   ~$&$~  0.4$  &  $ 5.10$&$0.32$  &  $7.55   ~$&$~  1.03$  &  3  \\
    \hline
  \end{tabular}
  \label{tab:mltbcs}
\end{table*}

\begin{table*}
  \centering
  \caption{Redshifts and derived properties of REFLEX clusters from follow-up observations (see caption for \tabref{tab:mltbcs}).}
  \begin{tabular}{|llr@{$\pm$}lr@{$\pm$}lr@{$\pm$}lr@{$\pm$}lr@{$~\pm~$}lr@{$\pm$}lc|}
    \hline
    Name & \multicolumn{1}{c}{$z$} & \multicolumn{2}{c}{$r_{500}$} & \multicolumn{2}{c}{$\Mgas{}_{,500}$} & \multicolumn{2}{c}{$M_{500}$} & \multicolumn{2}{c}{$L_{500}$} & \multicolumn{2}{c}{$\mysub{L}{500,ce}$} & \multicolumn{2}{c}{$\mysub{kT}{ce}$} & $kT$ \\
    & & \multicolumn{2}{c}{(Mpc)} & \multicolumn{2}{c}{$(10^{14}\Msun)$} & \multicolumn{2}{c}{$(10^{14}\Msun)$} & \multicolumn{2}{c}{$(10^{44}\erg\second^{-1})$} & \multicolumn{2}{c}{$(10^{44}\erg\second^{-1})$} & \multicolumn{2}{c}{(keV)} &  ref \\
    \hline
Abell 3558         &  0.048   &  $1.28  ~$&$~  0.10$  &  $0.73  ~$&$~  0.09$  &  $6.4   ~$&$~  1.0$  &  $\hspace{2.5ex}3.7   ~$&$~  0.3$  &  $ 2.60$&$0.24$  &  $5.51   ~$&$~  0.10$  &  1  \\
Abell 85           &  0.0557  &  $1.33  ~$&$~  0.06$  &  $0.82  ~$&$~  0.09$  &  $7.2   ~$&$~  1.0$  &  $5.7   ~$&$~  0.4$  &  $ 3.04$&$0.20$  &  $6.45   ~$&$~  0.10$  &  2  \\
Abell 3667         &  0.0557  &  $1.57  ~$&$~  0.10$  &  $1.35  ~$&$~  0.21$  &  $11.8  ~$&$~  2.2$  &  $5.8   ~$&$~  0.5$  &  $ 4.38$&$0.40$  &  $6.33   ~$&$~  0.06$  &  2  \\
Abell 3266         &  0.0602  &  $1.45  ~$&$~  0.07$  &  $1.06  ~$&$~  0.13$  &  $9.2   ~$&$~  1.4$  &  $4.9   ~$&$~  0.3$  &  $ 3.77$&$0.24$  &  $8.63   ~$&$~  0.18$  &  2  \\
Abell 3112         &  0.0752  &  $1.10  ~$&$~  0.05$  &  $0.47  ~$&$~  0.06$  &  $4.1   ~$&$~  0.6$  &  $4.4   ~$&$~  0.4$  &  $ 1.73$&$0.16$  &  $4.28   ~$&$~  0.09$  &  1  \\
Abell 2597         &  0.0852  &  $0.97  ~$&$~  0.06$  &  $0.33  ~$&$~  0.05$  &  $2.9   ~$&$~  0.5$  &  $4.3   ~$&$~  0.4$  &  $ 1.29$&$0.13$  &  $3.58   ~$&$~  0.07$  &  1  \\
Abell 3921         &  0.094   &  $1.20  ~$&$~  0.09$  &  $0.62  ~$&$~  0.09$  &  $5.4   ~$&$~  0.9$  &  $3.1   ~$&$~  0.3$  &  $ 2.23$&$0.21$  &  $5.07   ~$&$~  0.17$  &  1  \\
MS J1111.8         &  0.1306  &  $1.12  ~$&$~  0.09$  &  $0.52  ~$&$~  0.07$  &  $4.5   ~$&$~  0.7$  &  $3.2   ~$&$~  0.3$  &  $ 2.29$&$0.22$  &  $5.79   ~$&$~  0.22$  &  1  \\
Abell 1689         &  0.1832  &  $1.45  ~$&$~  0.07$  &  $1.21  ~$&$~  0.14$  &  $10.5  ~$&$~  1.5$  &  $13.6  ~$&$~  1.2$  &  $ 5.72$&$0.57$  &  $9.15   ~$&$~  0.35$  &  1  \\
Abell 2163         &  0.203   &  $2.22  ~$&$~  0.10$  &  $4.40  ~$&$~  0.45$  &  $38.5  ~$&$~  5.0$  &  $28.7  ~$&$~  1.1$  &  $17.77$&$0.67$  &  $12.27  ~$&$~  0.90$  &  3  \\
Abell 209          &  0.206   &  $1.53  ~$&$~  0.08$  &  $1.44  ~$&$~  0.18$  &  $12.6  ~$&$~  1.9$  &  $8.6   ~$&$~  0.3$  &  $ 5.80$&$0.22$  &  $8.23   ~$&$~  0.66$  &  3  \\
RX J1504.1$-$0248  &  0.2153  &  $1.46  ~$&$~  0.06$  &  $1.25  ~$&$~  0.13$  &  $11.0  ~$&$~  1.4$  &  $27.6  ~$&$~  1.0$  &  $ 5.74$&$0.26$  &  $8.00   ~$&$~  0.44$  &  3  \\
RX J0304.1$-$3656  &  0.2192  &  $1.07  ~$&$~  0.07$  &  $0.51  ~$&$~  0.08$  &  $4.4   ~$&$~  0.9$  &  $3.0   ~$&$~  0.2$  &  $ 2.05$&$0.12$  &  $6.27   ~$&$~  0.76$  &  3  \\
RX J0237.4$-$2630  &  0.2216  &  $1.16  ~$&$~  0.08$  &  $0.65  ~$&$~  0.11$  &  $5.6   ~$&$~  1.1$  &  $5.6   ~$&$~  0.3$  &  $ 2.59$&$0.18$  &  $6.65   ~$&$~  1.26$  &  3  \\
Abell 2667         &  0.2264  &  $1.36  ~$&$~  0.07$  &  $1.03  ~$&$~  0.14$  &  $9.0   ~$&$~  1.5$  &  $12.6  ~$&$~  0.6$  &  $ 4.97$&$0.31$  &  $7.58   ~$&$~  0.73$  &  3  \\
RX J0638.7$-$5358  &  0.2266  &  $1.42  ~$&$~  0.06$  &  $1.18  ~$&$~  0.13$  &  $10.3  ~$&$~  1.4$  &  $11.2  ~$&$~  0.6$  &  $ 5.48$&$0.31$  &  $9.45   ~$&$~  0.95$  &  3  \\
RX J0220.9$-$3829  &  0.228   &  $1.07  ~$&$~  0.08$  &  $0.50  ~$&$~  0.11$  &  $4.4   ~$&$~  1.1$  &  $4.0   ~$&$~  0.2$  &  $ 2.09$&$0.13$  &  $5.17   ~$&$~  0.53$  &  3  \\
Abell 521          &  0.2475  &  $1.46  ~$&$~  0.07$  &  $1.31  ~$&$~  0.16$  &  $11.4  ~$&$~  1.7$  &  $7.1   ~$&$~  0.3$  &  $ 5.80$&$0.21$  &  $6.21   ~$&$~  0.28$  &  3  \\
RX J0307.0$-$2840  &  0.2537  &  $1.28  ~$&$~  0.07$  &  $0.90  ~$&$~  0.13$  &  $7.8   ~$&$~  1.4$  &  $7.3   ~$&$~  0.4$  &  $ 3.73$&$0.22$  &  $9.60   ~$&$~  1.49$  &  3  \\
RX J2011.3$-$5725  &  0.2786  &  $0.95  ~$&$~  0.07$  &  $0.37  ~$&$~  0.07$  &  $3.3   ~$&$~  0.7$  &  $3.8   ~$&$~  0.2$  &  $ 1.67$&$0.12$  &  $3.23   ~$&$~  0.34$  &  3  \\
RX J0232.2$-$4420  &  0.2836  &  $1.49  ~$&$~  0.10$  &  $1.45  ~$&$~  0.25$  &  $12.7  ~$&$~  2.5$  &  $13.3  ~$&$~  0.7$  &  $ 6.57$&$0.44$  &  $10.06  ~$&$~  2.31$  &  3  \\
RX J0528.9$-$3927  &  0.2839  &  $1.52  ~$&$~  0.07$  &  $1.52  ~$&$~  0.17$  &  $13.3  ~$&$~  1.8$  &  $11.6  ~$&$~  0.6$  &  $ 6.92$&$0.39$  &  $7.80   ~$&$~  0.85$  &  3  \\
RX J0043.4$-$2037  &  0.2924  &  $1.28  ~$&$~  0.07$  &  $0.92  ~$&$~  0.14$  &  $8.1   ~$&$~  1.4$  &  $7.7   ~$&$~  0.4$  &  $ 4.75$&$0.29$  &  $7.59   ~$&$~  0.77$  &  3  \\
1ES 0657$-$558     &  0.2965  &  $1.81  ~$&$~  0.07$  &  $2.61  ~$&$~  0.24$  &  $22.8  ~$&$~  2.8$  &  $21.7  ~$&$~  0.3$  &  $13.61$&$0.26$  &  $11.70  ~$&$~  0.22$  &  3  \\
Abell 2537         &  0.2966  &  $1.23  ~$&$~  0.06$  &  $0.82  ~$&$~  0.11$  &  $7.2   ~$&$~  1.1$  &  $6.4   ~$&$~  0.4$  &  $ 3.87$&$0.29$  &  $7.63   ~$&$~  0.86$  &  3  \\
    \hline
  \end{tabular}
  \label{tab:mltreflex}
\end{table*}

\begin{table*}
  \centering
  \caption{Redshifts and derived properties of Bright MACS clusters from follow-up observations (see caption for \tabref{tab:mltbcs}). The last two entries represent clusters from \citet{Ebeling10} whose follow-up observations were not used in this paper, but which we include in this table for completeness; for these clusters we did not calculate the center-excised luminosity.}
  \begin{tabular}{|llr@{$\pm$}lr@{$\pm$}lr@{$\pm$}lr@{$\pm$}lr@{$~\pm~$}lr@{$\pm$}lc|}
    \hline
    Name & \multicolumn{1}{c}{$z$} & \multicolumn{2}{c}{$r_{500}$} & \multicolumn{2}{c}{$\Mgas{}_{,500}$} & \multicolumn{2}{c}{$M_{500}$} & \multicolumn{2}{c}{$L_{500}$} & \multicolumn{2}{c}{$\mysub{L}{500,ce}$} & \multicolumn{2}{c}{$\mysub{kT}{ce}$} & $kT$ \\
    & & \multicolumn{2}{c}{(Mpc)} & \multicolumn{2}{c}{$(10^{14}\Msun)$} & \multicolumn{2}{c}{$(10^{14}\Msun)$} & \multicolumn{2}{c}{$(10^{44}\erg\second^{-1})$} & \multicolumn{2}{c}{$(10^{44}\erg\second^{-1})$} & \multicolumn{2}{c}{(keV)} &  ref \\
    \hline
MACS J2245.0+2637   & 0.301  & $1.08 ~$&$~ 0.07$ & $0.56 ~$&$~ 0.10$ &  $4.9   ~$&$~  1.0$  &  $\hspace{2.5ex}7.6   ~$&$~  0.4$  &  $3.09$&$0.24$  &  $5.47   ~$&$~  0.58$  &  3  \\
MACS J1131.8$-$1955 & 0.306 & $1.69 ~$&$~ 0.09$ & $2.16 ~$&$~ 0.28$ &  $18.9  ~$&$~  2.9$  &  $13.1  ~$&$~  0.7$  &  $ 8.48$&$0.49$  &  $9.35   ~$&$~  1.67$  &  3  \\
MACS J0014.3$-$3022 & 0.308  & $1.65 ~$&$~ 0.07$ & $2.01 ~$&$~ 0.21$ &  $17.6  ~$&$~  2.3$  &  $13.6  ~$&$~  0.4$  &  $10.07$&$0.28$  &  $8.53   ~$&$~  0.37$  &  3  \\
MACS J2140.2$-$2339 & 0.313  & $1.06 ~$&$~ 0.04$ & $0.54 ~$&$~ 0.05$ &  $4.7   ~$&$~  0.6$  &  $11.1  ~$&$~  0.4$  &  $ 2.90$&$0.13$  &  $4.67   ~$&$~  0.43$  &  3  \\
MACS J0242.5$-$2132 & 0.314  & $1.25 ~$&$~ 0.07$ & $0.88 ~$&$~ 0.13$ &  $7.7   ~$&$~  1.3$  &  $14.2  ~$&$~  0.8$  &  $ 3.64$&$0.27$  &  $4.99   ~$&$~  0.77$  &  3  \\
MACS J1427.6$-$2521 & 0.318 & $0.94 ~$&$~ 0.06$ & $0.37 ~$&$~ 0.06$ &  $3.3   ~$&$~  0.6$  &  $4.1   ~$&$~  0.2$  &  $ 1.74$&$0.09$  &  $4.86   ~$&$~  0.57$  &  3  \\
MACS J0547.0$-$3904 & 0.319  & $1.02 ~$&$~ 0.07$ & $0.49 ~$&$~ 0.10$ &  $4.3   ~$&$~  1.1$  &  $6.4   ~$&$~  0.4$  &  $ 2.63$&$0.17$  &  $4.66   ~$&$~  0.53$  &  3  \\
MACS J0257.6$-$2209 & 0.322 & $1.23 ~$&$~ 0.06$ & $0.86 ~$&$~ 0.10$ &  $7.5   ~$&$~  1.1$  &  $7.0   ~$&$~  0.4$  &  $ 3.73$&$0.22$  &  $7.01   ~$&$~  0.88$  &  3  \\
MACS J2049.9$-$3217 & 0.323  & $1.18 ~$&$~ 0.06$ & $0.75 ~$&$~ 0.10$ &  $6.6   ~$&$~  1.0$  &  $6.1   ~$&$~  0.3$  &  $ 3.71$&$0.23$  &  $8.12   ~$&$~  1.18$  &  3  \\
MACS J2229.7$-$2755 & 0.324 & $1.11 ~$&$~ 0.06$ & $0.63 ~$&$~ 0.08$ &  $5.5   ~$&$~  0.8$  &  $10.0  ~$&$~  0.4$  &  $ 2.96$&$0.15$  &  $5.81   ~$&$~  0.68$  &  3  \\
MACS J1319.9+7003   & 0.327 & $1.06 ~$&$~ 0.06$ & $0.55 ~$&$~ 0.09$ &  $4.8   ~$&$~  0.9$  &  $4.2   ~$&$~  0.3$  &  $ 2.66$&$0.20$  &  $8.40   ~$&$~  2.41$  &  3  \\
MACS J0520.7$-$1328 & 0.336  & $1.16 ~$&$~ 0.06$ & $0.73 ~$&$~ 0.09$ &  $6.3   ~$&$~  0.9$  &  $7.9   ~$&$~  0.4$  &  $ 4.03$&$0.27$  &  $6.50   ~$&$~  0.84$  &  3  \\
MACS J1931.8$-$2634 & 0.352 & $1.34 ~$&$~ 0.07$ & $1.14 ~$&$~ 0.15$ &  $9.9   ~$&$~  1.6$  &  $19.7  ~$&$~  1.0$  &  $ 5.62$&$0.41$  &  $7.47   ~$&$~  1.40$  &  3  \\
MACS J0035.4$-$2015 & 0.352 & $1.35 ~$&$~ 0.07$ & $1.17 ~$&$~ 0.15$ &  $10.2  ~$&$~  1.6$  &  $11.9  ~$&$~  0.6$  &  $ 6.36$&$0.42$  &  $7.29   ~$&$~  0.66$  &  3  \\
MACS J0947.2+7623   & 0.354 & $1.27 ~$&$~ 0.07$ & $0.97 ~$&$~ 0.13$ &  $8.5   ~$&$~  1.3$  &  $20.0  ~$&$~  1.0$  &  $ 4.54$&$0.34$  &  $9.46   ~$&$~  2.14$  &  3  \\
MACS J1115.8+0129   & 0.355 & $1.28 ~$&$~ 0.06$ & $0.99 ~$&$~ 0.11$ &  $8.6   ~$&$~  1.2$  &  $14.5  ~$&$~  0.5$  &  $ 5.43$&$0.25$  &  $9.20   ~$&$~  0.98$  &  3  \\
MACS J0308.9+2645   & 0.356 & $1.58 ~$&$~ 0.08$ & $1.87 ~$&$~ 0.23$ &  $16.4  ~$&$~  2.4$  &  $14.7  ~$&$~  0.9$  &  $ 8.43$&$0.57$  &  $9.97   ~$&$~  1.05$  &  3  \\
MACS J0404.6+1109   & 0.352 & $1.17 ~$&$~ 0.11$ & $0.78 ~$&$~ 0.22$ &  $6.8   ~$&$~ 2.1$   &  $4.3  ~$&$~  0.6$   &  $ 3.69$&$0.45$  &  $7.66   ~$&$~  2.79$  &  3  \\
MACS J1532.8+3021   & 0.363 & $1.31 ~$&$~ 0.08$ & $1.08 ~$&$~ 0.17$ &  $9.5   ~$&$~  1.7$  &  $19.8  ~$&$~  0.7$  &  $ 5.16$&$0.27$  &  $6.83   ~$&$~  1.00$  &  3  \\
MACS J0011.7$-$1523 & 0.379 & $1.19 ~$&$~ 0.06$ & $0.83 ~$&$~ 0.10$ &  $7.2   ~$&$~  1.1$  &  $8.9   ~$&$~  0.3$  &  $ 4.25$&$0.19$  &  $6.80   ~$&$~  0.61$  &  3  \\
MACS J0949.8+1708   & 0.384 & $1.38 ~$&$~ 0.09$ & $1.30 ~$&$~ 0.22$ &  $11.3  ~$&$~  2.3$  &  $10.6  ~$&$~  0.6$  &  $ 6.23$&$0.41$  &  $8.92   ~$&$~  1.83$  &  3  \\
MACS J1720.2+3536   & 0.387 & $1.14 ~$&$~ 0.07$ & $0.72 ~$&$~ 0.11$ &  $6.3   ~$&$~  1.1$  &  $10.2  ~$&$~  0.4$  &  $ 4.31$&$0.20$  &  $7.90   ~$&$~  0.74$  &  3  \\
MACS J1731.6+2252   & 0.389 & $1.43 ~$&$~ 0.17$ & $1.47 ~$&$~ 0.25$ &  $12.8  ~$&$~  2.4$  &  $9.3   ~$&$~  0.5$  &  $ 7.42$&$0.43$  &  $5.87   ~$&$~  0.61$  &  3  \\
MACS J2211.7$-$0349 & 0.396 & $1.61 ~$&$~ 0.07$ & $2.06 ~$&$~ 0.23$ &  $18.1  ~$&$~  2.5$  &  $24.0  ~$&$~  1.2$  &  $10.15$&$0.63$  &  $13.97  ~$&$~  2.74$  &  3  \\
MACS J0429.6$-$0253 & 0.399 & $1.10 ~$&$~ 0.05$ & $0.66 ~$&$~ 0.08$ &  $5.8   ~$&$~  0.8$  &  $10.9  ~$&$~  0.6$  &  $ 3.91$&$0.25$  &  $8.33   ~$&$~  1.58$  &  3  \\
MACS J0159.8$-$0849 & 0.407 & $1.35 ~$&$~ 0.06$ & $1.23 ~$&$~ 0.13$ &  $10.8  ~$&$~  1.4$  &  $16.0  ~$&$~  0.6$  &  $ 6.53$&$0.28$  &  $9.11   ~$&$~  0.68$  &  3  \\
MACS J2228.5+2036   & 0.411 & $1.49 ~$&$~ 0.07$ & $1.68 ~$&$~ 0.21$ &  $14.7  ~$&$~  2.2$  &  $13.3  ~$&$~  0.7$  &  $ 8.63$&$0.49$  &  $7.36   ~$&$~  0.82$  &  3  \\
MACS J0152.5$-$2852 & 0.413 & $1.21 ~$&$~ 0.11$ & $0.90 ~$&$~ 0.23$ &  $7.9   ~$&$~  2.2$  &  $8.6   ~$&$~  0.5$  &  $ 5.02$&$0.37$  &  $4.71   ~$&$~  0.49$  &  3  \\
MACS J1206.2$-$0847 & 0.439 & $1.61 ~$&$~ 0.08$ & $2.19 ~$&$~ 0.29$ &  $19.2  ~$&$~  3.0$  &  $21.1  ~$&$~  1.1$  &  $10.55$&$0.64$  &  $10.71  ~$&$~  1.29$  &  3  \\
MACS J0417.5$-$1154 & 0.443 & $1.69 ~$&$~ 0.07$ & $2.53 ~$&$~ 0.25$ &  $22.1  ~$&$~  2.7$  &  $29.1  ~$&$~  1.5$  &  $15.29$&$0.94$  &  $9.49   ~$&$~  1.12$  &  3  \\
MACS J2243.3$-$0935 & 0.447 & $1.56 ~$&$~ 0.07$ & $1.98 ~$&$~ 0.24$ &  $17.4  ~$&$~  2.5$  &  $15.2  ~$&$~  0.8$  &  $11.56$&$0.67$  &  $8.24   ~$&$~  0.92$  &  3  \\
MACS J1347.5$-$1144 & 0.451 & $1.67 ~$&$~ 0.08$ & $2.48 ~$&$~ 0.27$ &  $21.7  ~$&$~  3.0$  &  $42.2  ~$&$~  1.1$  &  $11.47$&$0.35$  &  $10.75  ~$&$~  0.83$  &  3\vspace{2mm} \\
MACS J0358.8$-$2955 & 0.425 & $1.50 ~$&$~ 0.19$ & $1.81 ~$&$~ 0.47$ &  $15.8  ~$&$~  4.6$  &  $18.9  ~$&$~  1.2$  &  \multicolumn{2}{c}{}  &  $9.87  ~$&$~  1.21$  &  3  \\
MACS J2311.5+0338   & 0.305 & $1.64 ~$&$~ 0.13$ & $1.98 ~$&$~ 0.42$ &  $17.4  ~$&$~  4.1$  &  $12.9  ~$&$~  1.0$  &  \multicolumn{2}{c}{}  &  $8.25  ~$&$~  1.26$  &  3  \\
    \hline
  \end{tabular}
  \label{tab:mltmacs}
\end{table*}

\subsubsection{\ROSAT{} data analysis}
The \ROSAT{} PSPC observations were reduced using the Extended Source Analysis Software package of \citet{Snowden94}. Time intervals of high particle and scattered solar X-ray background were excluded, along with periods of detector instability. Remaining particle and solar X-ray backgrounds were modeled in the nominal energy bands of 0.7--0.9, 0.9--1.3, and 1.3--2.0\keV{} (standard \ROSAT{} bands R5--R7) and spatially distributed into background maps. Exposure maps were generated in each energy band from detector maps obtained during the RASS; these exposure maps take into account vignetting and detector artifacts. Flat-fielded images were created by subtracting the background maps from the data and correcting with the exposure maps. Cluster centroids were determined and surface brightness profiles were created, after masking all detectable non-cluster sources. Due to the limited energy coverage of the PSPC, \ROSAT{} observations are insufficient to determine temperatures for the hot clusters under study; instead, we adopt temperatures from the literature, measured using {\it ASCA} \citep{Horner01} or \Chandra{} \citep{Vikhlinin09} (\tabrefs{tab:mltbcs}--\ref{tab:mltreflex}). Assuming a typical metallicity, $Z=0.3$, we generated Monte Carlo samples of the \ROSAT{} count rate to flux conversion factor as above, utilizing response and effective area files for each observation generated for the appropriate region (the central ring) of the PSPC. For the high-temperature clusters under study, uncertainties in the temperatures and metallicities have a very small impact on the measured quantities of interest: luminosity, radius ($r_{500}$) and gas mass. Associated systematic uncertainties are well within the tolerances marginalized over in \secref{sec:deproj}.

\subsubsection{Gas mass, total mass, and luminosity measurements} \label{sec:deproj}
Our analysis uses gas mass as a robust, low-scatter proxy for total cluster mass. This choice is motivated by the fact that \Mgas{} can be measured robustly, independent of the dynamical state of the cluster; observations have also indicated consistency and minimal scatter in the gas mass fraction, $\fgas=\Mgas/\Mtot$, at $r_{2500}$ for hot, massive clusters \citep[][hereafter \citetalias{Allen08}]{Allen08}, in agreement with hydrodynamical simulations \citep*[e.g.][]{Nagai07}.\footnote{Note that, as detailed in \citetalias{Allen08}, \fgas{} can only be measured reliably for dynamically relaxed clusters, and is most accurately measured at radii $r \sim r_{2500}$. For dynamically active systems, large systematic variations in measured \fgas{} values are expected due to strong, time variable departures from hydrostatic equilibrium \citep{Nagai07}; the same is true of \fgas{} measurements at large radii, regardless of dynamical state. However, the true (as opposed to measured) \fgas{} values at $r \geq r_{2500}$ for all clusters are still expected to be similar; that is, mergers effectively disrupt hydrostatic equilibrium, but do not affect the gas mass fraction at these radii.} The value of \fgas{} at $r_{500}$ can be inferred from $\fgas(r_{2500})$ using simulated cluster \fgas{} profiles, accounting for appropriate systematic uncertainties, providing a simple conversion from \Mgas{} to \Mtot{} at $r_{500}$.

Deprojected gas mass profiles were generated for each cluster based on its 0.7--2.0\keV{} surface brightness profile, using a modified version of the Cambridge X-ray deprojection code of \citet*{White97}. Briefly, the gas mass profile is non-parametrically reconstructed using the surface brightness profile and a model for the total mass distribution, which is adjusted to reproduce the measured average temperature. As noted previously, the emissivity in the 0.7--2.0\keV{} band is very weakly dependent on temperature for the hot clusters in our sample, so the derived gas mass is largely insensitive to this procedure; typically, changes greater than a factor of two in the predicted temperature are required to shift the gas mass results by $1\sigma$. Uncertainties on the gas mass profiles were determined by repeating this procedure for many Monte Carlo realizations of the surface brightness profiles.

Hydrodynamical simulations indicate that systematic biases in gas mass measurement due to effects such as asphericity, clumping and projection are small for the general cluster population at all radii of interest. Motivated by the simulations of \citet{Nagai07}, we incorporate a systematic fractional bias (and associated uncertainty) in gas mass measurements at $r_{500}$: $B=0.0325\pm 0.06$ (consistent with zero). With this systematic allowance, we expect that the gas mass values determined by our procedure to be robust.

The gas masses are converted to total masses using the measured gas mass fraction of massive clusters. \citetalias{Allen08} showed that, for hot, massive clusters, the ratio $\fgas(r_{2500})$ is consistent with a universal value \citep*[see also simulations of, e.g.,][]{Eke98a,Nagai07}. Given a model for the gas mass fraction profile, the gas mass profiles can be used to determine total masses, $M_\Delta$, and associated radii, $r_\Delta$, corresponding to mean enclosed overdensity $\Delta$, by solving the implicit equation
\begin{equation}
  \label{eq:rDelta}
  M(r_\Delta) = \frac{\Mgas(r_\Delta)}{(1+B) \fgas(r_\Delta)} = \frac{4\pi\Delta\rhoc(z) r_\Delta^3}{3(1+B)},
\end{equation}
where $\rhoc(z)$ is the critical density at the redshift of the cluster, for $r_\Delta$ and $M(r_\Delta)$. Adopting the value $\fgas(r_{2500})=0.1104$ from \citetalias{Allen08}, we verified that our analysis produces estimates of $r_{2500}$ and $M_{2500}$ that are compatible with the more detailed results of that work for the systems in common.

To obtain mass estimates at larger radii ($\Delta=500$) from the gas mass profiles, we need an estimate of \fgas{} at this radius. The increased prevalence of departures from hydrostatic equilibrium at these larger radii \citep[e.g.][]{Nagai07} prevent us from measuring $\fgas(r_{500})$ directly, even for dynamically relaxed systems. Instead, we use the prediction of simulations \citep[e.g.][]{Eke98a,Crain07} that hot, massive clusters should display approximately self-similar \fgas{} profiles at radii $r \geq r_{2500}$ (approximately one quarter of the virial radius), independent of their dynamical state. Using the simulations of \citet{Eke98a}, we fit a power-law model to $\fgas(r)$ between $r_{2500}$ and $r_{500}$, finding a logarithmic slope of $0.048 \pm 0.030$. Assuming the total mass follows a \citet*{Navarro97} profile with a typical concentration parameter $c=4$ \citep{Zhao03,Zhao08,Gao08}, the ratio $r_{500}/r_{2500}$ is approximately 2.3, yielding a scaled, mean gas mass fraction estimate of $\fgas(r_{500})\approx 0.115$ from the \citetalias{Allen08} value of $\fgas(r_{2500})$. We note that this procedure is insensitive to the assumed concentration parameter, since $(r_{500}/r_{2500})^{0.048}$ is a weak function of $c$, varying at only the per cent level between $c=2$ and $c=10$. The systematic uncertainty associated with the concentration is thus small in comparison with the other allowances described below, as well as the statistical uncertainties.

The dominant systematic uncertainty affecting our total mass measurements is associated with the measurement $\fgas(r_{2500})$. For the purposes of obtaining mass estimates at $r_{500}$, we adopted $\fgas(r_{2500})=0.1104$ as a reference value, similar to the cosmological reference values $h=0.7$ and $\Omegam=0.3$. The systematic uncertainty on the universal \fgas{} value is fully accounted for in the analysis by incorporating and marginalizing over the detailed model of \citetalias{Allen08} (including systematic allowances at the 10--20~per cent level; see \secref{sec:fgasdata}) and modeling the dependence of measured quantities on \fgas{} in a similar manner to the cosmological reference parameters (see \cosmopaper{}). Systematic, cluster-to-cluster scatter in \fgas{} is undetected in the sample of large, dynamically relaxed clusters studied in \citetalias{Allen08}; we nevertheless incorporated an additional systematic uncertainty in the individual mass estimates at the 5 per cent level to account both for possible intrinsic scatter and residual calibration or modeling uncertainties. This uncertainty is reflected in the error bars listed in \tabrefs{tab:mltbcs}--\ref{tab:mltmacs}.

Luminosities in the RASS band (0.1--2.4\keV{}) within $r_{500}$ were calculated by integrating the surface brightness profiles, using the count rate to flux conversions computed from the spectral analysis. This integration was accomplished by fitting the profile outskirts with a power-law cluster emission plus constant background model; the luminosities were background subtracted and, where necessary, the power-law model was used to extrapolate into the low signal-to-noise regime. Uncertainties from the profile modeling were propagated by Monte Carlo sampling parameters of the fits, and were combined with a 5 per cent systematic allowance accounting for residual observation-to-observation calibration uncertainties. The mass and luminosity measurements appear in \tabrefs{tab:mltbcs}--\ref{tab:mltmacs}.

We also calculated projected luminosities for each cluster within the annulus $0.15<r/r_{500}<1$. The procedure is the same as above, except that the integration of the surface brightness profile begins at $r=0.15r_{500}$ rather than $r=0$. We note that, for the $z<0.2$ clusters where we use \ROSAT{} data, the resolution of the PSPC is sufficient to perform this center-excision. The resulting luminosities, denoted $\mysub{L}{500,ce}$, are investigated in \secref{sec:mlce}, but do not contribute to the main results of this paper, or the cosmological constraints in \cosmopaper{}.

\subsubsection{Instrument cross-calibration} \label{sec:crosscal}
There are 11 clusters in our sample for which both \ROSAT{} and \Chandra{} follow-up observations were analyzed (see \tabrefs{tab:bcsobs}--\ref{tab:macsobs}). These clusters were used to cross-calibrate the two instruments, although in the end only the \Chandra{} data for these 11 clusters were used to determine the properties listed in \tabrefs{tab:mltbcs}--\ref{tab:mltmacs}. For the purposes of the cross-calibration, the masses and luminosities from \ROSAT{} were estimated using centroids, temperatures and metallicities determined from \Chandra{} and the comparisons were made at a fixed radius, the \Chandra-determined value of $r_{500}$.

Using {\sc ciao} version 4.1.1 and CALDB 4.1.2, the luminosities determined from \Chandra{} data are systematically larger than those from \ROSAT{} by 14 per cent, with a scatter of 7 per cent. This disagreement is predominantly a consequence of the 21 January 2009 \Chandra{} calibration update, which significantly changed the ACIS effective area at soft energies. A previous analysis using CALDB version 3.4.3 and CIAO 3.4 (which also did not incorporate corrections for dead area in the front-illuminated chips caused by cosmic rays, now accounted for by default) produced near perfect agreement between \Chandra{} and \ROSAT{} luminosities. Since our ultimate goal is to relate cluster masses to \ROSAT{} survey fluxes, we apply this 14 per cent correction to the \Chandra{} luminosities in \tabrefs{tab:mltbcs}--\ref{tab:mltmacs}. Our results on the luminosity--mass and temperature--luminosity relations are thus for \ROSAT-calibrated luminosities; \Chandra-calibrated relations can be derived simply by adjusting the normalizations of the relations (\secrefs{sec:model} and \ref{sec:results}) appropriately.

Our \Chandra{} gas mass estimates, which to good approximation scale as the square root of the luminosity, were correspondingly larger by $\sim 7$ per cent, with a scatter of 2 per cent. We again calibrated the \Chandra{} gas masses to the \ROSAT{} standard, which agrees extremely well with previous versions of the \Chandra{} calibration. In this case, the correction is motivated by the need for consistency between gas masses measured in this study with those of \citetalias{Allen08}, which predate the change to the ACIS effective area. In taking the ratio $\fgas/\Mgas$ (\eqnref{eq:rDelta}), with \fgas{} from \citetalias{Allen08} and \Mgas{} measured in this study, any common systematic bias in gas mass cancels, leaving our estimates of the total masses unbiased. We note that the 21 January 2009 \Chandra{} calibration update additionally affects temperature measurements, which is relevant for the \citetalias{Allen08} determination of \fgas{}; we address this issue in \secref{sec:fgasdata}.

\subsection{\fgas{} data} \label{sec:fgasdata}

Since our analysis uses gas mass as a proxy for total cluster mass, we must simultaneously constrain both the gas mass fraction and the growth of structure. For that reason, our standard analysis includes the \fgas{} analysis of \citetalias{Allen08}; however, we use only the 6 lowest-redshift clusters from that paper ($z<0.15$). Those 6 clusters are sufficient to constrain the gas mass fraction at low redshift, without themselves producing a direct constraint on the expansion of the Universe (and thus unduly influencing our cosmological results). This simultaneous analysis incorporates the full \fgas{} model detailed in \citetalias{Allen08}, and includes generous systematic allowances for instrument calibration (10 per cent), non-thermal pressure support \citep[10 per cent,][]{Nagai07}, the depletion of baryons in clusters with respect to the cosmic mean (20 per cent), and evolution with redshift of the baryonic and stellar content of clusters (10 and 20 per cent).

The raw data analysis in \citetalias{Allen08} predates a recent (21 January 2009) update to the \Chandra{} ACIS effective area at soft energies. As discussed in \secref{sec:crosscal}, the effect of the calibration on the \citetalias{Allen08} gas masses can be compensated by calibrating our own gas masses consistently. However, the calibration update has an additional effect on the determination of \fgas{}; the change in measured temperatures due to the relative effective area at low and high energies results in a corresponding change in total mass estimates. In our tests, this effect results in an increase in \fgas{} values for massive clusters at the 5--10~per cent level; we have accounted for this correction by shifting the center of the ``\Chandra{} calibration'' nuisance parameter in the \citetalias{Allen08} model such that the preferred value of \fgas{} increases correspondingly by 10 per cent. The Gaussian prior on this parameter has itself a width of 10 per cent, so the systematic allowance encompasses both the old value and the new value expected from the calibration update.

\subsection{Other data} \label{sec:otherdata}

As discussed above, our analysis of the XLF data includes 6 low redshift ($z<0.15$) \fgas{} clusters from \citetalias{Allen08}. As in \citetalias{Mantz08}, when analyzing the X-ray cluster data alone, we also incorporate Gaussian priors on the Hubble constant, $h=0.72 \pm 0.08$, based on the Hubble Key project \citep{Freedman01}, and on the mean baryon density, $\Omegab h^2=0.0214 \pm 0.002$, based on big bang nucleosynthesis studies \citep{Kirkman03}. These priors are not needed when cosmic microwave background (CMB) data are also included in the analysis; see below.

In \secref{sec:results}, we present results from a combination of XLF and other independent cosmological data, including cluster gas mass fractions \citepalias[][including the full 42 cluster data set]{Allen08}, 5-year CMB data from the {\it Wilkinson Microwave Anisotropy Probe} \citep[WMAP5,][]{Dunkley09}, the type Ia supernova (SNIa) analysis of \citet[][including their treatment of systematic errors]{Kowalski08}, and the Baryon Acoustic Oscillation (BAO) analysis of \citet{Percival07}. More details on these data and their use are reviewed in \cosmopaper{}.

When fitting CMB data, we allow the scalar spectral index, $\mysub{n}{s}$, and optical depth to reionization, $\tau$, to vary as free parameters, and marginalize over a plausible range in the amplitude of the Sunyaev-Zel'dovich signal due to galaxy clusters ($0<\mysub{A}{SZ}<2$; introduced by \citealt{Spergel07}). The combination of CMB and \fgas{} data places tight constraints on both $h$ and $\Omegab h^2$ in addition to other parameters of interest (see \citetalias{Allen08}), so the Hubble Key project and big bang nucleosynthesis priors are unnecessary in analyses of the combined data sets. This applies as well to the combination of CMB and XLF data, since we always use a low-$z$ subset of the \fgas{} data to calibrate the cluster mass scale for the XLF analysis.

\section{Scaling relation model} \label{sec:model}

The model used for the joint cosmology and scaling relation work is described in full in \cosmopaper{}. Here we review the basic scaling relation model and describe various possible extensions to it that are investigated in \secref{sec:extensions}.

In \cosmopaper{}, the nominal values of luminosity and temperature for a given mass are modeled as power laws:
\begin{eqnarray}
  \label{eq:nominalMLT}
  \expectation{\ell(m)} &=& \beta_0^{\ell m} + \beta_1^{\ell m} m, \\
  \expectation{t(m)} &=& \beta_0^{tm} + \beta_1^{tm} m. \nonumber
\end{eqnarray}
Here the terms
\begin{eqnarray}
  \label{eq:MLTdefs}
  \ell &=& \logTen\left(\frac{L_{500}}{E(z)10^{44}\erg\second^{-1}}\right), \nonumber\\
  m &=& \logTen\left(\frac{E(z)M_{500}}{10^{15}\Msun}\right), \\
  t &=& \logTen\left(\frac{\mysub{kT}{ce}}{\keV}\right), \nonumber
\end{eqnarray}
include factors of the normalized Hubble parameter, $E(z)=H(z)/H_0$, expected for ``self-similar'' evolution of the relations with redshift. These factors appear because the relation is defined in terms of quantities measured within radius $r_{500}$, and follow from the dependence of $r_{500}$ on the evolving critical density of the Universe \citep[e.g.][]{Kaiser86,Bryan98}.\footnote{We note that evolution of this form is a generic prediction of gravitationally driven collapse, appearing ubiquitously in simulations \citep[e.g.][]{Evrard08}, and does not require the assumption of virial equilibrium.} The intrinsic scatter in $\ell$ and $t$ at fixed $m$ is non-negligible, and can be modeled most simply as bivariate normal, described by marginal scatters, $\sigma_{\ell m}$ and $\sigma_{tm}$, and a scatter correlation coefficient, $\rho_{\ell t m}$. We refer to this seven parameter model ($\beta_0^{\ell m}$, $\beta_1^{\ell m}$, $\beta_0^{tm}$, $\beta_1^{tm}$, $\sigma_{\ell m}$, $\sigma_{tm}$, $\rho_{\ell t m}$) as the ``minimal'' scaling relation model.

This model can be naturally extended in several ways. One possibility is to allow departures from self-similar evolution in the nominal relations. We write
\begin{eqnarray}
  \label{eq:nominalMLT2}
  \expectation{\ell(m)} &=& \beta_0^{\ell m} + \beta_1^{\ell m} m + \beta_2^{\ell m}\logTen(\zeta), \\
  \expectation{t(m)} &=& \beta_0^{tm} + \beta_1^{tm} m + \beta_2^{tm}\logTen(\zeta), \nonumber
\end{eqnarray}
where $\zeta$ is a function of redshift, commonly chosen to be $1+z$ (e.g. \citealt{Ettori04}; \citealt*{Morandi07}; \citetalias{Mantz08}) or $E(z)$ \citep[e.g.][]{Maughan07,Vikhlinin09}.

We similarly consider the possibility of evolution in the intrinsic scatter, making each scatter parameter ($\sigma_{\ell m}$, $\sigma_{tm}$ and $\rho_{\ell t m}$) a function of redshift of the form
\begin{equation}
  \label{eq:MLTscatev}
  \sigma_{\ell m}(z) = \sigma_{\ell m} \left[ 1 + \sigma_{\ell m}' (\zeta-1) \right].
\end{equation}

A third potentially interesting avenue is to allow asymmetry in the intrinsic scatter. We consider generalizing the bivariate normal distribution used to describe the scatter to the bivariate skew-normal distribution \citep[for reference, see][]{Azzalini85,Azzalini96,Azzalini99}. For simplicity, we only consider skewness in the marginal luminosity--mass relation, parametrized by the shape parameter $\lambda_{\ell m}$.

\section{Analysis} \label{sec:analysis}

The analysis method used in this paper is described in full in \cosmopaper{}. Uniquely, it provides a rigorous and fully self-consistent means of producing simultaneous constraints on both the cluster scaling relations and cosmological parameters; by this we mean that the likelihood of the full data set (survey + follow-up observations) is derived from first principles, ensuring that the covariance among all the model parameters is fully captured and that selection biases\footnote{Since we are concerned with selection biases relative to a mass-selected sample, the intrinsic dispersion between X-ray luminosity and mass is an important (indeed, the dominant) contributing factor, in addition to survey flux measurement errors (see also Appendix~\ref{sec:pedbias}).} (Malmquist bias and Eddington bias) are properly accounted for. The results are determined by sampling of the posterior distribution using Markov Chain Monte Carlo, marginalizing over priors encoding all appropriate systematic uncertainties (see \cosmopaper{}). The most important aspects in this context are that (1) the sample selection functions\footnote{By ``selection function'', we refer to the probability that a cluster be detected by a survey and included in the data set, as a function of that cluster's physical properties (redshift, flux etc.). See Section~4.1 of \cosmopaper{}.} are included; (2) the cluster mass function is taken into account; and (3) the correlation of measurement errors is properly incorporated into the likelihood. As reviewed in Appendix~\ref{sec:pedbias}, the first two points above are crucial for a proper accounting of Malmquist and Eddington biases.\footnote{While the mass function clearly must be taken into account to properly handle Eddington and Malmquist biases (see Appendix~\ref{sec:pedbias}), there is a second, independent effect for which it is relevant. Generically, but especially in the presence of intrinsic scatter, the distribution of covariates (in this case $m$) is significant. For example, in methods based on least squares, the data at the ends of the range of covariates are disproportionately influential in determining the slope; thus, even in schemes that are ostensibly modified to include intrinsic scatter \citep[e.g.][]{Press92}, the assumption of uniformly distributed covariates can produce biased results (see \citealt{Kelly07} and references therein).} The above factors, as well as the large size and high completeness and purity of the cluster samples, and the quality of the follow-up data, lead to significant improvements over previous work.

Consistent with the method described above, all 238 clusters meeting our selection criteria are used in the analysis presented here. However, the 94 clusters with follow-up data naturally have more influence on the scaling relation results (for more details, see \cosmopaper{}, Section~4.1.3).

In this paper, the only cosmological model we consider is the simple, spatially flat, cosmological constant model (\LCDM{}) which is known to provide a good fit to CMB and kinematic data \citep[e.g. \citetalias{Allen08};][]{Kowalski08,Dunkley09}. Our results for the cluster scaling relations are marginalized over the parameters of this cosmological model, which are enumerated in \cosmopaper{}. Constraints on cosmology from our data are also presented in that work.

\section{Results on the scaling relations} \label{sec:results}

\subsection{Constraints on the minimal model} \label{sec:minmodres}

\subsubsection{The joint luminosity--temperature--mass relations} \label{sec:mainres}
\begin{figure*}
  \centering
  \includegraphics[scale=0.7]{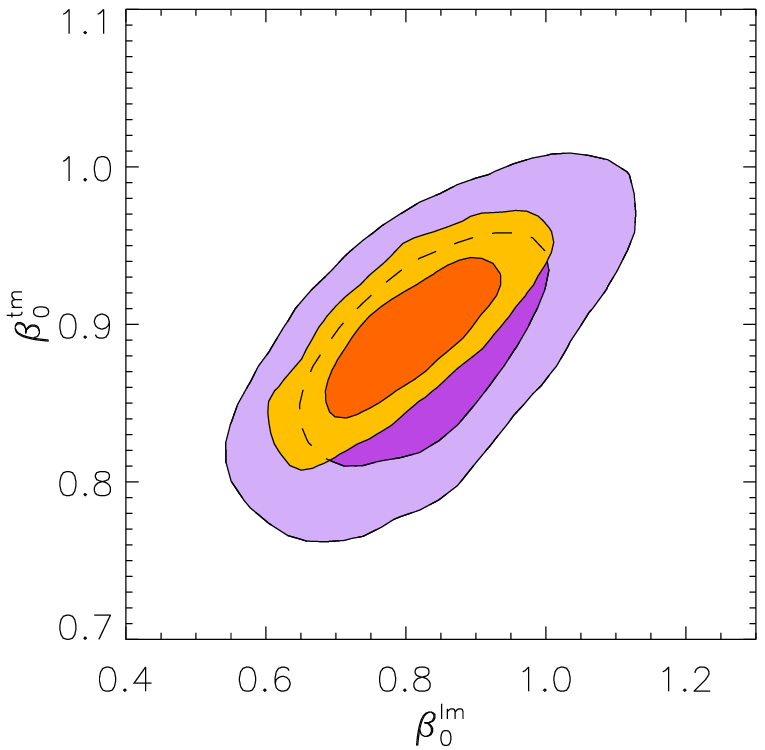}
  \hspace{0.25cm}
  \includegraphics[scale=0.7]{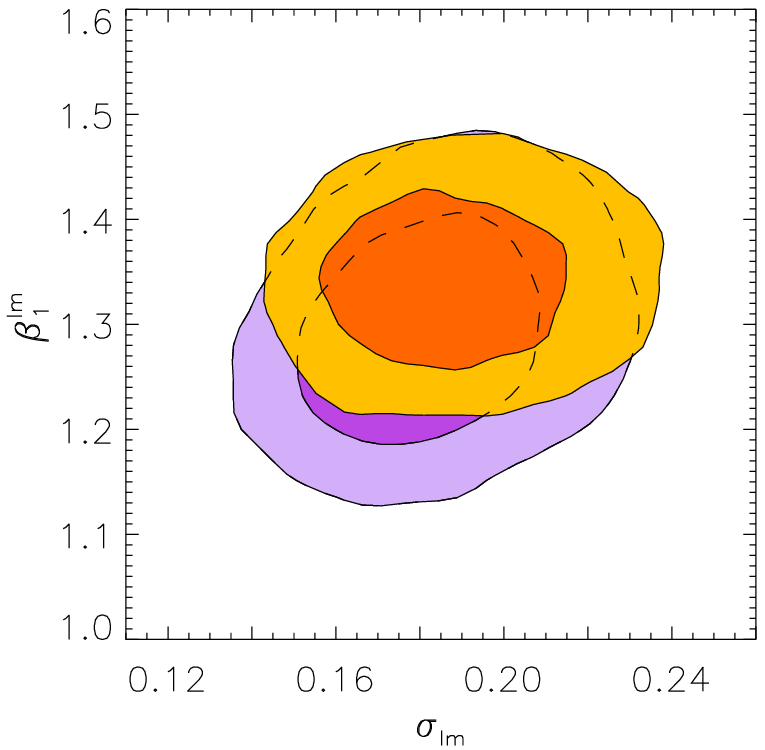}
  \hspace{0.25cm}
  \includegraphics[scale=0.7]{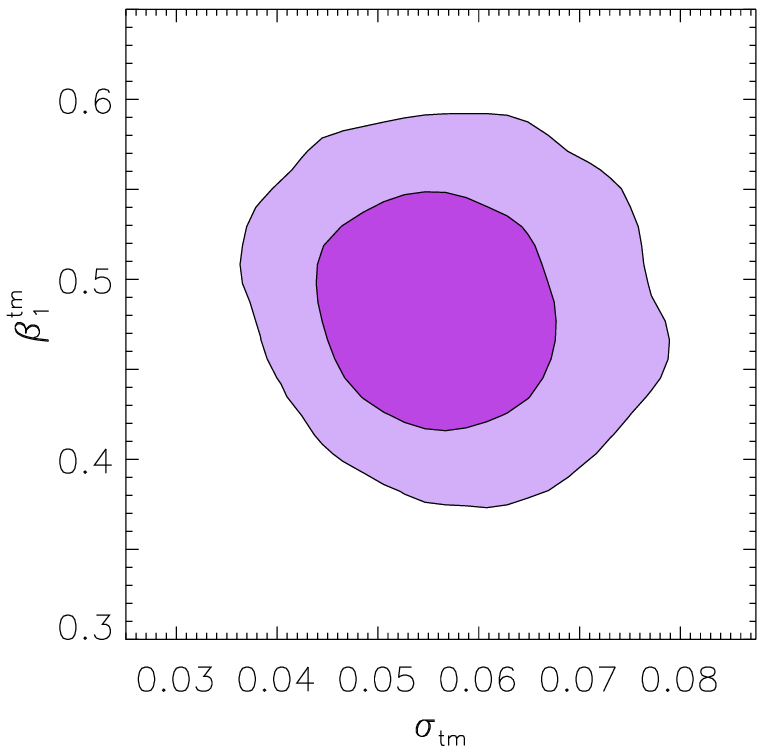}
  \caption{Joint 68.3 and 95.4 per cent confidence regions on parameters of the minimal (7 parameter) scaling relation model from XLF data (purple) and from the combination with independent cosmological data (gold, \secref{sec:otherdata}). Left: normalizations of the nominal luminosity--mass and temperature--mass relations. The degeneracy is principally due to the fact that both parameters correlate with the overall mass scale set by \fgas{}. Center: slope and marginal intrinsic scatter of the luminosity--mass relation. Right: same for the temperature--mass relation. The combination with other data is not shown in this panel, since the results are very similar (\tabref{tab:minmodres}).}
  \label{fig:minmodres}
\end{figure*}

One-dimensional, marginalized constraints on the parameters of the minimal scaling relation model from the XLF data are presented in \tabref{tab:minmodres}. Joint constraints on the most degenerate parameter pairs are shown in \figref{fig:minmodres}; the only significant degeneracy for this simple model is between the normalizations of the nominal luminosity--mass and temperature--mass relations, both of which strongly correlate with \fgas{} and cosmological parameters.\footnote{The normalization and slope of each nominal relation are not strongly degenerate because we have balanced the data around an appropriate pivot ($10^{15}\Msun$; \eqnref{eq:MLTdefs}).}

\begin{table}
  \centering
  \caption{Marginalized 68.3 per cent confidence intervals on the parameters of the minimal (7 parameter) scaling relation model from XLF data alone, and from the combination of all the cosmological data sets (\secref{sec:otherdata}). Tildes indicate parameters describing the bolometric luminosity--mass (as opposed to 0.1--2.4\keV{} luminosity) relation; the bolometric results are derived from the 0.1--2.4\keV{} results using the temperature--mass relation. $\mysub{t}{ci}$ indicates temperature measured in the full $r<r_{500}$ region of each cluster (without excising the center).}
  \label{tab:minmodres}
  \begin{tabular}{l@{\hspace{2mm}}c@{\hspace{2mm}}cc}
    \hline
    Parameter & & XLF only & all data \\
    \hline
    $\ell$--$m$ normalization              & $\beta_0^{\ell m}$ & $0.82 \pm 0.11$ & $0.80 \pm 0.08$\vspace{0.5mm} \\
    $\ell$--$m$ slope                      & $\beta_1^{\ell m}$ & $1.29 \pm 0.07$ & $1.34 \pm 0.05$\vspace{0.5mm} \\
    $t$--$m$ normalization                 & $\beta_0^{tm}$     & $0.88 \pm 0.05$ & $0.89 \pm 0.03$\vspace{0.5mm} \\
    $t$--$m$ slope                         & $\beta_1^{tm}$     & $0.48 \pm 0.04$   & $0.49 \pm 0.04$\vspace{0.5mm} \\
    $\ell$--$m$ marginal scatter           & $\sigma_{\ell m}$  & $0.180 \pm 0.019$ & $0.185 \pm 0.019$\vspace{0.5mm} \\
    $t$--$m$ marginal scatter              & $\sigma_{tm}$      & $0.056 \pm 0.008$ & $0.055 \pm 0.008$\vspace{0.5mm} \\
    scatter correlation                    & $\rho_{\ell t m}$  & $0.10 \pm 0.19$ & $0.09 \pm 0.19$\tablespace \\
    $\mysub{\ell}{bol}$--$m$ normalization & $\tilde{\beta}_0^{\ell m}$ & $1.26 \pm 0.17$ & $1.23 \pm 0.12$\vspace{0.5mm} \\
    $\mysub{\ell}{bol}$--$m$ slope         & $\tilde{\beta}_1^{\ell m}$ & $1.59 \pm 0.09$ & $1.63 \pm 0.06$\tablespace \\
    $\mysub{t}{ci}$--$m$ normalization     & $\beta_{0\mathrm{,ci}}^{tm}$ & $0.90 \pm 0.04$   & $0.91 \pm 0.03$\vspace{0.5mm} \\
    $\mysub{t}{ci}$--$m$ slope             & $\beta_{1\mathrm{,ci}}^{tm}$ & $0.46 \pm 0.04$   & $0.46 \pm 0.04$\vspace{0.5mm} \\
    $\mysub{t}{ci}$--$m$ marginal scatter  & $\sigma^\mathrm{ci}_{tm}$    & $0.059 \pm 0.007$ & $0.060 \pm 0.07$\vspace{0.5mm} \\
    \hline
  \end{tabular}
\end{table}

\begin{figure*}
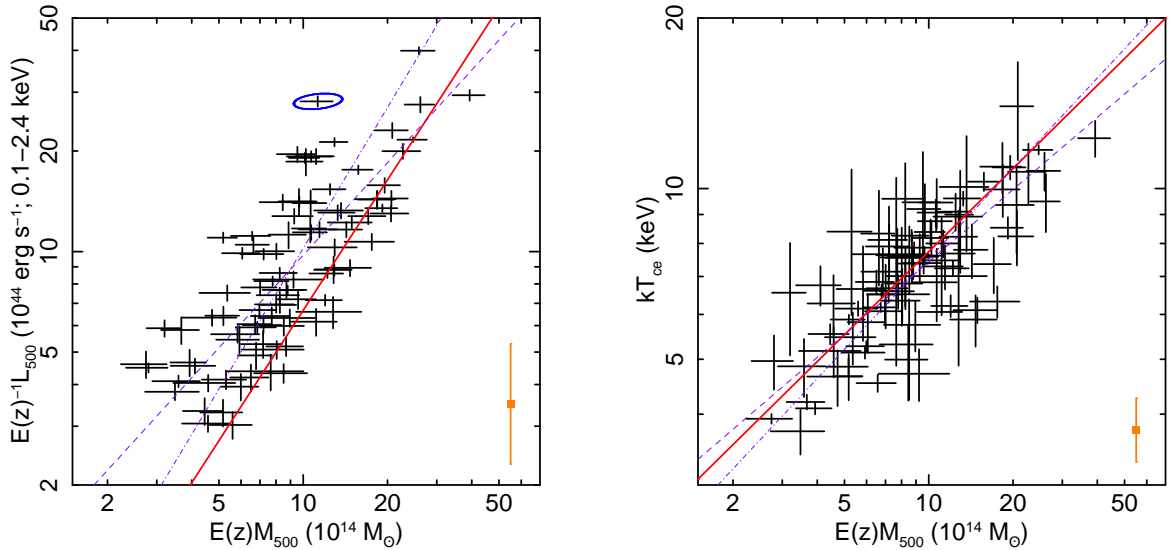

  \centering
  \includegraphics[scale=0.925]{fig3a.eps}
  \hspace{1cm}
  \includegraphics[scale=0.925]{fig3b.eps}
  \caption{Mass, luminosity and temperature data for our cluster sample. Note that the data are adjusted with factors of $E(z)$ appropriate for the best-fitting cosmology (\cosmopaper), not the reference cosmology. Also shown are the best-fitting, nominal luminosity--mass and temperature--mass power-law relations (solid, red lines), and the results of fits to the data with the BCES $Y|X$ and $X|Y$ methods (purple lines, respectively dashed and dot-dashed). The best-fitting values of the intrinsic scatter in each relation, $\sigma_{\ell m}$ and $\sigma_{tm}$, are indicated by the orange points with error bars in the lower-right corner of each panel. The seemingly low normalization of our luminosity--mass fit compared with the data reflects the effects of Malmquist and Eddington biases on the sample, which are not accounted for in the BCES fits. Plotted error bars do not include the contributions due to systematic uncertainty in \fgas{}, or uncertainty in the overall instrument calibration, although these are included in the analysis. The correlation between luminosities and masses measured from the same X-ray data are accounted for (see \cosmopaper{}); this small correlation ($\rho \approx 0.3$) is illustrated by the blue ellipse in the left panel, which shows the (joint, 68.3 per cent confidence) error ellipse for one cluster. There is no correlation between measured temperatures and masses, using our procedure (\secref{sec:pointed}).}
  \label{fig:fitMLT}
\end{figure*}

The luminosity--mass and temperature--mass data and the best-fitting nominal relations from the analysis are shown in \figref{fig:fitMLT}. For the purposes of this visual comparison, the data have been adjusted by the $E(z)$ factors appropriate to the best-fitting cosmology and to account for the cosmological dependence of $r_{500}$ (a smaller correction; see Section 3.4 of \cosmopaper{}). For the luminosity--mass relation in the 0.1--2.4\keV{} band, we measure normalization $\beta_0^{\ell m}=0.82 \pm 0.11$, slope $\beta_1^{\ell m}=1.29 \pm 0.07$, and scatter $\sigma_{\ell m}=0.180 \pm 0.019$ ($\sim 40$ per cent).

The apparent offset in normalization of the best fit (red line) from the luminosity--mass data reflects the dramatic effect of Malmquist and Eddington biases on the data set. Specifically, the flux limits of our cluster samples are high enough that the data set consists largely of sources that are of average luminosity or greater, given their masses. \citet{Pacaud07} observed a similar effect in an analysis of {\it XMM-Newton} Large Scale Structure ({\it XMM}-LSS) clusters that accounted for the survey selection function, as did \citet{Stanek06}, who incorporated both the selection function and cluster mass function into an analysis of the REFLEX sample. These results underscore the fact that selection effects must be fully accounted for in order to accurately recover the scaling relations from cluster data.

In contrast, naively applying the commonly used BCES $Y|X$ or $X|Y$ methods of \citet{Akritas96} to the plotted data results in the purple dashed and dot-dashed lines in \figref{fig:fitMLT}, respectively. For the luminosity--mass relation (left panel), both BCES fits, which do not account for selection bias, have slopes and normalizations that are inconsistent with the true best fit (red, solid line). The discrepancy is significant for cosmological applications, since the luminosity--mass relation is influential in determining the total number of cluster detections, and is thus degenerate with \Omegam{} and $\sigma_8$. We have similarly fit the data in \figref{fig:fitMLT} with other simple regression methods based on modified least-squares techniques \citep[the {\sc fitexy} routine of][]{Press92} and Bayes theorem\footnote{We refer specifically to the {\sc linmix\_err} routine, which does not account for selection bias. Since our own analysis is also Bayesian, it can be thought of as a specialization of the general approach advocated by \citet{Kelly07}.} \citep{Kelly07}, and find that they produce results very similar to the BCES $Y|X$ and $X|Y$ methods, depending on which variable is assigned to be the covariate.

For the temperature--mass relation (right panel of \figref{fig:fitMLT}; using center-excised temperatures, $\mysub{kT}{ce}$), we find normalization $\beta_0^{tm}=0.88 \pm 0.05$, slope $\beta_1^{tm}=0.48 \pm 0.04$, and scatter $\sigma_{tm}=0.056 \pm 0.008$ ($\sim 13$ per cent scatter). Since this relation is not as influenced by selection bias as the luminosity--mass relation is, the BCES fits are much closer to our best fit in normalization in this case.

\begin{figure}
  \centering
  \includegraphics[scale=0.9]{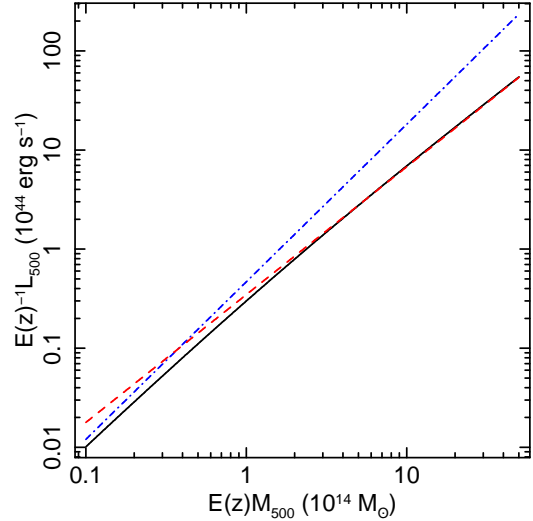}
  \caption{Best-fitting luminosity--mass relations, where luminosity is either bolometric or in the 0.1--2.4\keV{} band. The dot-dashed, blue line and dashed, red line show respectively the best-fitting bolometric and band luminosity relations, extrapolated to lower masses. The solid, black line shows the predicted band luminosity relation, assuming the best-fitting bolometric relation is valid at all masses and using the best-fitting temperature--mass relation. If the true, fundamental bolometric relation is a simple power law, a break in the slope of the band luminosity relation is expected at $\sim 10^{14}\Msun$; however, the band luminosity relation is still well described by a power law for $M>2\E{14}\Msun$.}
  \label{fig:LbolLband}
\end{figure}

The slope $\beta_1^{\ell m}$ can be simply converted to a bolometric luminosity--mass slope, $\tilde{\beta}_1^{\ell m}$, using the temperature--mass relation. (We use a tilde to distinguish parameters describing the bolometric luminosity relation.) The low scatter of the temperature--mass relation, together with the fact that the ratio of 0.1--2.4\keV{} band luminosity to bolometric luminosity is weakly dependent on redshift and close to linear in temperature (for $kT>1\keV$), means that a power law in mass and band luminosity implies a power law in mass and bolometric luminosity (and vice versa) to good approximation over the mass range of our data ($>3\E{14}\Msun$; see \figref{fig:LbolLband}). Performing this conversion, our best-fitting band luminosity--mass relation corresponds to a bolometric luminosity--mass relation with normalization $\tilde{\beta}_0^{\ell m}=1.26 \pm 0.17$ and slope $\tilde{\beta}_1^{\ell m}=1.59 \pm 0.09$.

Finally, we note that our temperature--mass relation, determined with temperatures measured in the (0.15--1)$r_{500}$ annulus, is not significantly different from that obtained using temperatures measured within the full region $r<r_{500}$. For this center-included (ci) temperature--mass relation, we find normalization $\beta_{0\mathrm{,ci}}^{tm}=0.90 \pm 0.04$, slope $\beta_{1\mathrm{,ci}}^{tm}=0.46 \pm 0.04$, and marginal scatter $\sigma^\mathrm{ci}_{tm}=0.059 \pm 0.007$ from the analysis of the XLF data.

\subsubsection{Results incorporating external cosmological data}
The precision of our results can be improved somewhat by incorporating the external cosmological data sets summarized in \secref{sec:otherdata}. These additional data reduce the posterior uncertainty on the cosmological parameters of the model, allowing the XLF data to more effectively constrain the scaling relations. (Recall that the parameters describing the scaling relations and the underlying cosmology are determined simultaneously.) The results of this combined analysis are shown in \tabrefs{tab:minmodres} and \ref{tab:ltres}, and in \figref{fig:minmodres}. Improvements are limited to the normalizations of the two relations, $\beta_0^{\ell m}$ and $\beta_0^{tm}$, and the luminosity--mass slope, $\beta_1^{\ell m}$. The former are primarily degenerate with \fgas{}, which sets the overall mass scale, while the latter is degenerate with $\sigma_8$ and influences the total number of clusters that should be detected.

\subsubsection{The derived temperature--luminosity relation}

\begin{table}
  \centering
  \caption{Marginalized 68.3 per cent confidence intervals on the parameters of the temperature--luminosity relation from XLF data alone, and from the combination of all the cosmological data sets (\secref{sec:otherdata}). Tildes indicate parameters describing a bolometric (as opposed to 0.1--2.4\keV{} band) luminosity relation, which are derived from the 0.1--2.4\keV{} results using the temperature--mass relation (\tabref{tab:minmodres}).}
  \label{tab:ltres}
  \begin{tabular}{lccc}
    \hline
    Parameter & & XLF only & all data \\
    \hline
    $t$--$\ell$ normalization              & $\beta_0^{t\ell}$ & $0.55 \pm 0.08$ & $0.57 \pm 0.07$\vspace{0.5mm} \\
    $t$--$\ell$ slope                      & $\beta_1^{t\ell}$ & $0.30 \pm 0.08$ & $0.29 \pm 0.08$\vspace{0.5mm} \\
    $t$--$\ell$ scatter                    & $\sigma_{t\ell}$  & $0.08 \pm 0.02$ & $0.08 \pm 0.02$\tablespace \\
    $t$--$\mysub{\ell}{bol}$ normalization & $\tilde{\beta}_0^{t\ell}$ & $0.48 \pm 0.08$ & $0.50 \pm 0.07$\vspace{0.5mm} \\
    $t$--$\mysub{\ell}{bol}$ slope         & $\tilde{\beta}_1^{t\ell}$ & $0.27 \pm 0.04$ & $0.27 \pm 0.04$ \\
    \hline
  \end{tabular}
\end{table}

Given the model defined in \secref{sec:model}, the nominal temperature--luminosity relation and its scatter are not trivially related to the fitted parameters, even at $z=0$. The stochastic relation is
\begin{equation}
  \label{eq:LTrel}
  P(t|\ell) = \frac{P(\ell,t)}{P(\ell)} = \frac{\int dm P(\ell,t|m)P(m)}{\int dm P(\ell|m)P(m)},
\end{equation}
where $P(\ell,t|m)$ is the bivariate scaling relation, $P(\ell|m)$ is the marginal luminosity--mass relation, and $P(m)$ is proportional to the mass function. Thus, a detailed prediction for the temperature--luminosity relation requires specifying a set of cosmological parameters.\footnote{We note that the common practice of, e.g., estimating the temperature--luminosity slope as $\beta_1^{tm}/\beta_1^{\ell m}$ corresponds to the case of a flat mass function, and is therefore biased.} Defining the nominal relation
\begin{equation}
  \label{eq:LTnom}
  \expectation{t(\ell)} = \beta_0^{t\ell} + \beta_1^{t\ell}\ell
\end{equation}
with normal intrinsic scatter $\sigma_{t\ell}$, we find $\beta_0^{t\ell}=0.55 \pm 0.08$, $\beta_1^{t\ell}=0.30 \pm 0.08$ and $\sigma_{t\ell}=0.08 \pm 0.02$. Since the dependence of these estimates on the details of the mass function is not very strong, we have not determined these best fitting values and uncertainties precisely by marginalizing over cosmological parameters, although this would be straightforward in principle; instead, we have taken the less computationally intensive approach of using the mass function for our reference cosmology at $z=0$, and including in the error bars the difference between estimates based on the corresponding reference $P(m)$ and a uniform $P(m)$ as a conservative systematic allowance. These results, and the corresponding constraints for the bolometric temperature--luminosity relation appear in \tabref{tab:ltres}.

\subsubsection{The derived $Y_X$--mass relation}
The quantity $Y_X=\Mgas \mysub{kT}{ce}$, an approximation to the thermal energy in the intracluster medium, has been proposed as a low scatter proxy for total mass \citep*{Kravtsov06}, and has found application in cosmological work \citep{Vikhlinin09,Vikhlinin09a}. Using this definition, and the expectation that gas mass and total mass are directly proportional ($\Mgas=\fgas M$, with \fgas{} constant), the virial theorem predicts the $Y_X$--mass relation to be (cf. \eqnref{eq:virialpr})
\begin{equation}
  \label{eq:yx}
  Y_X \propto M \left[ E(z) M \right]^{2/3}.
\end{equation}
Defining
\begin{equation}
  \label{eq:ydef}
  y = \logTen \left( \frac{E(z)\Mgas{}(r_{500})\mysub{kT}{ce}}{10^{15}\Msun\keV} \right),
\end{equation}
and allowing a general temperature--mass slope, $\beta_1^{tm}$, we can write the self-similarly evolving $Y_X$--mass relation in terms of the gas mass fraction and the parameters of the temperature--mass relation:
\begin{eqnarray}
  \label{eq:YMrel}
  y &=& \beta_0^{ym} + \beta_1^{ym} m \\
  &=& \beta_0^{tm} + \logTen\left[ \fgas(r_{500}) \right] + \left( \beta_1^{tm} + 1\right) m. \nonumber
\end{eqnarray}
The best fit temperature--mass relation from the XLF data and mean \fgas{} value at $r_{500}$ thus predict a normalization of $\beta_0^{ym}=0.01 \pm 0.05$ and a slope of $\beta_1^{ym}=1.48 \pm 0.04$ for the $Y_X$--mass relation (\tabref{tab:YM}).

\begin{table}
  \centering
  \caption{Marginalized 68.3 per cent confidence intervals on the parameters of the $Y_X$--mass relation, derived from the best-fitting temperature--mass relation and mean \fgas{} value. The intrinsic scatter listed here is estimated from the scatter of the data about the derived relation, and is consistent with the value expected based on the scatter in the temperature--mass relation.}
  \label{tab:YM}
  \begin{tabular}{lcc}
    \hline
    Parameter & & Constraint \\
    \hline
    $y$--$m$ normalization      & $\beta_0^{ym}$ & $0.01 \pm 0.05$\vspace{0.5mm} \\
    $y$--$m$ slope              & $\beta_1^{ym}$ & $1.48 \pm 0.04$\vspace{0.5mm} \\
    $y$--$m$ scatter            & $\sigma_{ym}$  & $0.052 \pm 0.008$ \\
    \hline
  \end{tabular}
\end{table}

\begin{figure}
  \centering
  \includegraphics[scale=0.9]{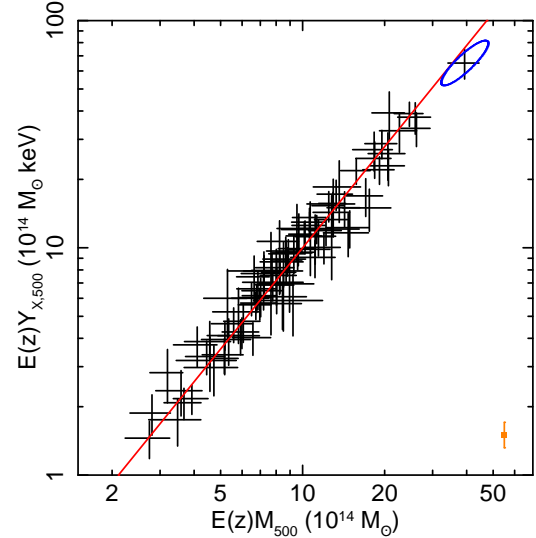}
  \caption{Total mass and $Y_X=\Mgas \mysub{kT}{ce}$ are plotted for our best-fitting cosmology (\cosmopaper) and compared with the $Y_X$--mass relation predicted by the best-fitting temperature--mass relation and \fgas{} value. The intrinsic scatter, indicated by the orange point with error bars, is similar to that of the temperature--mass relation ($\sim 0.05$). Note that the X and Y error bars are correlated, since \Mgas{} is used to compute both $Y_X$ and the total mass, $M_{500}$. To illustrate this correlation ($\rho \approx 0.85$), the (joint, 68.3 per cent confidence) error ellipse is shown in blue for the rightmost cluster.}
  \label{fig:YXfig}
\end{figure}

This prediction is compared with the data in \figref{fig:YXfig}. The scatter about the line is similar to the intrinsic scatter in the temperature--mass relation ($\sim 0.05$, or 12 per cent), as expected from \eqnref{eq:YMrel} and the fact that intrinsic scatter in \fgas{} for massive clusters remains undetected in current data (with a 68.3 per cent confidence upper bound $<7$ per cent; \citetalias{Allen08}). These considerations, and the fact that the gas mass can be measured both more precisely and with significantly smaller systematic uncertainties than temperature\footnote{Uncertainties in background modeling are currently the largest source of systematic error in temperature measurements at $r_{500}$.} for a given exposure time, suggest that \Mgas{} may be a preferable mass proxy to either $kT$ or $Y_X$, at least for the most massive clusters where \fgas{} is constant.

\subsection{Is the minimal model sufficient?} \label{sec:gof}

An important question to address is whether the minimal model, as we have defined it, provides an acceptable description of the data. Given the complexity and Bayesian nature of the model, testing the acceptability of the fit is necessarily more involved than following a simple $\chi^2$ prescription. Our approach, which is not a unique solution to the problem, is to compare each of the following aspects of each cluster sample to the predictions of the best fit:
\begin{enumerate}
\item the number of clusters detected,
\item the two-dimensional distribution of clusters in redshift and survey flux,
\item the marginal distribution of luminosities from follow-up observations given the mass and redshift measurements,
\item the marginal distribution of temperatures from follow-up observations given the measurements of mass, luminosity and redshift.
\end{enumerate}

The number of clusters in each sample is straightforwardly Poisson distributed. For each sample, we define a significance value as the probability of finding a number of clusters that is at least as discrepant from the predicted mean as the measured value.

To test the survey flux-redshift distribution of sources, we use a two-dimensional analog of the Kolmogorov-Smirnov (K-S) test due to \citet{Peacock83} and \citet[][see also \citealt{Press92}]{Fasano87}. Briefly, the test involves comparing the predicted and detected numbers of clusters in each of the four quadrants ($z\geq\hat{z},F\geq\hat{F}$; $z<\hat{z},F\geq\hat{F}$; etc.) about each cluster detection, $(\hat{z},\hat{F})$. The maximum discrepancy between these predictions and measurements is compared with a modification of the Kolmogorov $D$ distribution to obtain the associated significance.

The goodness of fit of the marginal luminosity--mass relation is tested by comparing the luminosities measured in the follow-up observations, $\hat{\ell}$, to their predicted distribution given the measured masses and redshifts, $\hat{m}$ and $\hat{z}$, and the detection of the cluster, indicated by $I$. This density, $P(\hat{\ell}'|\hat{z},\hat{m},I)$, is closely related to the probability associated with a detected cluster in the likelihood function (see Appendix~\ref{sec:plzmi}), the principle difference being that the survey flux measurement should not be taken into account here. That is, the predicted distribution of $\hat{\ell}$ accounts for the mass function, the measured mass with its statistical error, and the selection bias. However, it should not account for the specific flux value measured in the survey, or the expression would reduce to a test of how well the two flux measurements agree. This distribution can be calculated numerically as a function of $\hat{\ell}'$ and compared with the measured values, $\hat{\ell}$. Since the shape of $P(\hat{\ell}'|\hat{z},\hat{m},I)$ is slightly different for each cluster, these residuals are not individually statistically meaningful. However, the cumulative values $C=\int_0^{\hat{\ell}}P(\hat{\ell}'|\hat{z},\hat{m},I)d\hat{\ell}'$ will be distributed uniformly on $[0,1]$, provided that the predicted distributions are accurate, independent of the fact that those distributions are not identical. Thus, we can test whether the fit is a consistent description of the luminosity--mass relation by checking whether these $C$ values are uniformly distributed, using a one-sample K-S test. An analogous procedure is used to test the residuals of the temperature--mass relation.

For the purpose of this goodness of fit test, we use the set of parameters producing the highest-likelihood sample in the Markov chains; this need not be precisely the mode of the posterior distribution, but should be adequate for the purpose of testing the fit. The 12 significance values (4 tests for each of the 3 cluster samples) obtained through the procedure described here are displayed in \tabref{tab:gof}. None of the individual tests produced a sufficiently low significance value ($<0.01$) to indicate an inadequacy in the model. We note that if, for example, one test had produced a very small significance, it would still need to be interpreted in the context of the larger goodness of fit test; that is, the likelihood of randomly obtaining one low significance value out of 12 would need to be taken into account.

\begin{table}
  \centering
  \caption{Significance values for each goodness of fit test applied to each cluster sample. Very low values ($<0.01$) would indicate that the model is inadequate to describe the data (though, formally, the probability of randomly obtaining some low values in a sample of 12 would need to be taken into account).}
  \label{tab:gof}
  \begin{tabular}{lccc}
    \hline
    test & BCS & REFLEX & MACS \\
    \hline
    number & 0.73 & 0.69 & 0.21 \\
    redshift-flux & 0.18 & 0.78 & 0.79 \\
    luminosity--mass & 0.42 & 0.93 & 0.35 \\
    temperature--mass & 0.19 & 0.28 & 0.92 \\
    \hline
  \end{tabular}
\end{table}

\subsection{Extensions to the model} \label{sec:extensions}

\begin{table*}
  \centering
  \caption{Parameters which can extend the scaling relation model. Constraints of the form $X \pm \sigma_X$ are 68.3 per cent confidence intervals; the bound on $\lambda_{\ell m}$ refers to a sharp cut-off in posterior density. The notation $\mathcal{U}(a,b)$ indicates the uniform distribution with endpoints $a$ and $b$. Deviance information criteria are with respect to the best fit using the minimal scaling relation model.}
  \begin{tabular}{lclcl}
    \hline
    Parameter & & Prior & Constraint & $\Delta$DIC \\
    \hline
    Extra evolution in $\ell$--$m$ normalization & $\beta_2^{\ell m}$ & $\mathcal{U}(-2,2)$   & $-0.5 \pm 0.4$ & $-0.9$ \\
    Extra evolution in $t$--$m$ normalization    & $\beta_2^{tm}$     & $\mathcal{U}(-2,2)$   & $-0.2 \pm 0.2$ & $-1.6$ \\
    Evolution in $\ell$--$m$ marginal scatter    & $\sigma_{\ell m}'$ & $\mathcal{U}(-2,2)$   & $-0.6 \pm 0.3$ & $-1.2$ \\
    Evolution in $t$--$m$ marginal scatter       & $\sigma_{tm}'$     & $\mathcal{U}(-2,2)$   & none & \\
    Evolution in $\ell$--$t$ scatter correlation & $\rho_{\ell t m}'$ & $\mathcal{U}(-2,2)$   & none & \\
    Asymmetry in $\ell$--$m$ marginal scatter    & $\lambda_{\ell m}$ & $\mathcal{U}(-10,10)$ & $>-5$ & \\
    \hline
  \end{tabular}
  \label{tab:sclextpar}
\end{table*}

Although the minimal set of parameters produces an acceptable fit, we can still investigate whether any of the extensions to the simple model, summarized in \tabref{tab:sclextpar}, are preferred by the data. To do this, we have performed additional analyses with each of the extension parameters individually free. The additional data described in \secref{sec:otherdata} were included in these analyses in order to constrain the cosmology as much as possible, maximizing our sensitivity to the scaling relations. For the parameters controlling evolution with redshift, we choose the particular form of evolution $\zeta(z)=1+z$ in \eqnrefs{eq:nominalMLT2} and \ref{eq:MLTscatev}; however, our conclusions are identical if the other commonly chosen function, $\zeta(z)=E(z)$, is used.

The parameters controlling evolution in the marginal temperature--mass scatter and the scatter correlation coefficient, $\sigma_{tm}'$ and $\rho_{\ell t m}'$, were not constrained within the allowed region (priors in \tabref{tab:sclextpar}). The best-fitting temperature--mass scatter, $\sigma_{tm}=0.055$, is significantly smaller than the average measurement uncertainty on $t$, $\sim 0.1$, so the inability of the data to constrain evolution in the scatter is perhaps not surprising. Similarly, constraining the scatter correlation as a function of redshift would require many more data points, so that the correlation in $\ell$ and $t$ at similar masses could be estimated at multiple redshifts. The relatively large uncertainty on $\rho_{\ell t m}$ ($\pm 0.19$) indicates that this is a challenging measurement to make with the current data set, even without allowing evolution with redshift.

The posterior distribution of the shape parameter of the luminosity--mass marginal scatter was sharply bounded at $\lambda_{\ell m}>-5$, but was otherwise flat within the allowed region. Our inability to significantly constrain asymmetry in the luminosity--mass scatter is consistent with the idea that the current data consist predominantly of clusters that are at least of average brightness given their masses. That is, the data do not sample well the lower tail of the distribution of luminosity given mass; hence, we can only rule out very negative skewness, where scatter to higher luminosities is suppressed beyond what is observed.

The remaining parameters, governing the evolution of the normalizations of the luminosity--mass and temperature--mass relations and of the luminosity--mass marginal scatter ($\beta_2^{\ell m}$, $\beta_2^{tm}$ and $\sigma_{\ell m}'$), are well constrained; the 68.3 per cent confidence intervals obtained for each are listed in \tabref{tab:sclextpar}. The fact that these constraints exclude zero at the 1--2$\sigma$ level is not itself sufficient to conclude that the data require additional parameters in the model. Instead, a model selection test must be employed to determine whether the observed improvement in likelihood is significant compared to that expected to occur randomly when a new parameter is added. We address this question using the Deviance Information Criterion (DIC) of \citet[][see also \citealt{Liddle07}]{Spiegelhalter02}. This information criterion is well suited to Bayesian problems and, unlike the Bayesian evidence, does not depend strongly on the somewhat arbitrary width of the priors. The DIC is
\begin{equation}
  \label{eq:DICdef}
  \mathrm{DIC} = 2\ln\tilde{\like} - 2\overline{\ln\like}
\end{equation}
where \like{} is the likelihood, the bar indicates an average over the posterior, and $\tilde{\like}$ is the likelihood evaluated at some measure of center of the posterior such as the mean or mode. We use the maximum likelihood sampled by the Markov chains as $\tilde{\like}$.

The difference in DIC obtained for these three tests compared with that from the minimal model is listed in \tabref{tab:sclextpar}. The $\Delta$DIC values range from $-0.9$ to $-1.6$, which, according to the Jeffreys' scale conventionally used to interpret these values, indicates that none of the extensions to the model is strongly favored.\footnote{Typically, a $\Delta$DIC of $<-5$ is interpreted as ``strong'' evidence for the more complex model, and $<-10$ is ``conclusive.''} We conclude that the minimal scaling relation model provides an adequate description of the data, and conversely that the data do not prefer more complex models at a statistically significant level.

\subsection{Comparison with other work} \label{sec:compare}

Previous results on the slopes and scatters of the scaling relations studied here have varied widely. Since no other authors have accounted fully for the underlying cluster mass function and the sample selection function, as we do here, such large systematic variations could be expected to result from the use of different data sets and fitting methods. Moreover, our analysis uses \Mgas{} as a proxy for total mass, which should be more reliable in the mass range studied here than the hydrostatic mass estimates used in most previous work. This is particularly true for the temperature--mass relation, since hydrostatic mass measurements are highly correlated with measured temperatures. In addition, the 21 January 2009 \Chandra{} calibration update may be significant; the reduction of relative effective area at high energies disproportionally reduces temperatures of hot clusters compared with cool systems, resulting in a net flattening of the temperature--mass relation. We note also the possibility that the slopes of the scaling relations are a function of mass, in which case our results should only be compared with those of similarly high-mass cluster samples. Nevertheless, for completeness, we compare in this section our constraints with other recent results in the literature.

For the luminosity--mass relation, we find a slope of $\beta_1^{\ell m}=1.29 \pm 0.07$ for 0.1--2.4\keV{} luminosities, or $1.59 \pm 0.09$ for bolometric luminosities. This slope is steeper than that expected from the virial theorem ($4/3$ for the bolometric slope), but shallower than most other results, which typically span the range 1.5--1.8 for band luminosities or 1.7--2.3 for bolometric luminosities (e.g. \citealt{Reiprich02,Zhang07}; \citetalias{Mantz08}; \citealt{Rykoff08,Zhang08,Pratt09,Vikhlinin09}), though we note good agreement with a previous study of very luminous clusters \citep{Allen03}.

Our temperature--mass slope, $\beta_1^{tm}=0.48 \pm 0.04$, is shallower than the virial prediction of $2/3$ and most others in the literature \citep[typically 0.6--0.7, e.g.][]{Allen01,Finoguenov01,Arnaud05,Popesso05,Vikhlinin06,Sun09}, though there do exist some results with similar slopes \citep[e.g.][]{Ettori04}. As noted above, some of the differences may be due to our use of gas mass as a proxy for total mass, which should be more reliable than hydrostatic masses for studying the temperature--mass relation. Consistently, our estimate of the $Y_X$--mass slope is also somewhat shallower than that derived in other studies \citep[e.g.][]{Arnaud07,Sun09,Vikhlinin09}.

Some authors studying a combination of low and high redshift clusters have remarked that the evolution expected from self similarity appears empirically justified, in that low and high redshift clusters with the appropriate $E(z)$ factors appear to lie about the same line \citep[e.g.][]{Maughan08,Zhang08}. \citet{Ettori04} and \citet{Morandi07} went further, by adopting a model for evolution as in \eqnref{eq:nominalMLT2} with $\zeta=1+z$. Using a modified least-squares fit, they found no evidence for evolution beyond the self-similar expectation, in agreement with our conclusions.

\citet{Pratt09} and \citet{Vikhlinin09} found luminosity--mass (0.1--2.4\keV{} band) slopes $\sim 1.6$ using a procedure in which the luminosity data are corrected for the expected Malmquist bias, given a flux limit or full selection function. \citet{Vikhlinin09} also considered the possibility of departures from self-similar evolution, finding a marginal preference for non-self-similar evolution of the luminosity--mass relation. Their estimate of the intrinsic scatter in the relation agrees well with our own.

\section{Center-excised luminosity scaling relations} \label{sec:mlce}

\begin{figure}
  \centering
  \includegraphics[scale=0.9]{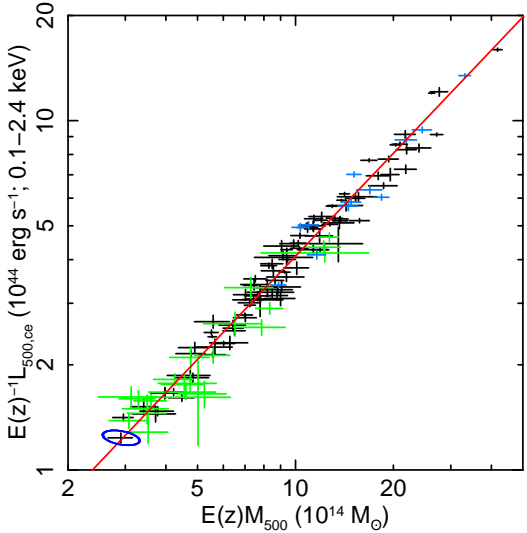}
  \caption{Luminosity--mass data for our reference cosmology, where luminosities include only emission from $0.15<r/r_{500}<1$. Note that the X and Y error bars are anti-correlated ($\rho \approx -0.4$), as illustrated by the blue (joint, 68.3 per cent confidence) error ellipse shown for the leftmost cluster; this anti-correlation is accounted for in the analysis. Black points indicate data used in our standard analysis; blue points are high-redshift ($0.5<z<0.7$) Faint MACS clusters from \citet{Ebeling07}, and green points are 400d clusters \citep[$0.35<z<0.9$;][]{Burenin07}. The best fit to this center-excised relation, shown in red, has a slope in agreement with the virial theorem and is consistent with having intrinsic scatter at the few per cent level. The data are consistent with self-similar evolution.}
  \label{fig:Lcice}
\end{figure}

It has long been known that the luminosity of clusters harboring cooling cores is significantly enhanced by the emission from bright, cool gas in their centers \citep[e.g.][]{Fabian94,Allen98a,Markevitch98,Peres98}. More recently, a dramatic reduction in luminosity--mass scatter has been demonstrated when luminosities are measured excluding cluster centers \citep[typically $r<0.15r_{500}$;][]{Maughan07,Zhang07}. Our data confirm and reinforce the conclusion that the center-excised luminosity--mass relation is significantly tighter than the center-included relation, as \figref{fig:Lcice} shows. (Note that, for clarity, error bars in the figure do not include systematic uncertainties that are used elsewhere in our analysis; see below.)

To the extent that the center-excised relation exhibits zero or negligible intrinsic scatter, selection effects will not bias the results of regression performed on these data. [However, it is still important to marginalize over cosmological parameters when fitting the data, since they are reflected in the values of $E(z)$.] If selection bias is not an issue, the strict requirements used to define the data set can be relaxed; \figref{fig:Lcice} shows the standard data set used throughout this work in black, with an additional 12 high-redshift ($0.5<z<0.7$) Faint MACS clusters from \citet{Ebeling07} in blue, and 22 clusters from the 400 Square Degree survey \citep[400d, $0.35<z<0.9$;][]{Burenin07} in green. The high-redshift MACS and 400d data satisfy \eqnref{eq:flimfcn} and the respective flux limits $10^{-12}\erg\second^{-1}\cm^{-2}$ (0.1--2.4\keV{} band) and $1.4\E{-13}\erg\second^{-1}\cm^{-2}$ (0.5--2.0\keV{}).\footnote{For the 400d clusters, we also require the core-included 0.1--2.4\keV{} luminosity from follow-up observations to be $>2\E{44}\erg\second^{-1}$, as the few clusters that otherwise meet our selection criteria have extremely poor signal-to-noise. The high-redshift MACS clusters are all above this luminosity limit.} \Chandra{} data for these clusters were reduced as described in \secref{sec:data}; details and results of the observations appear in \tabrefs{tab:moreobs} and \ref{tab:mltmore}.

We fit the relation
\begin{equation}
  \label{eq:MLce}
  \expectation{\mysub{\ell}{ce}(m)} = \gamma_0^{\ell m} + \gamma_1^{\ell m} m + \gamma_2^{\ell m}\logTen\left(\zeta\right),
\end{equation}
where $\mysub{\ell}{ce}$ is defined as in \eqnref{eq:MLTdefs}, but with luminosity from within radius $0.15r_{500}$ of the center excluded (see \secref{sec:deproj}). The fit was performed using the Bayesian linear mixture method of \citet{Kelly07}, which accounts for the correlation in mass and luminosity measurement errors.\footnote{We note that it is extremely straightforward, within the Bayesian framework, to account for correlation in the mass and luminosity measurements. In brief, the likelihood function integrates over the true values of mass and luminosity for every cluster, weighted by the sampling distribution, which gives the probability of obtaining the measured values given the true values. The sampling distribution in this calculation can easily be bivariate, with non-zero correlation. See also \citet{Gelman04}.} Somewhat counterintuitively, the correlation between center-excised luminosity and mass measurements is negative; although both quantities depend on the measured electron density, a stronger anti-correlation arises because a larger mass implies a larger $r_{500}$, which results in a net reduction in luminosity due to the center-excision.

\begin{table}
  \centering
  \caption{Marginalized 68.3 per cent confidence intervals on power law models describing the center-excised luminosity--mass and temperature--luminosity relations from our standard and extended data sets (see text). $\gamma_0$ and $\gamma_1$ are the normalization and slope of each relation when luminosities are measured in the 0.1--2.4\keV{} band;  $\tilde{\gamma}_0$ and $\tilde{\gamma}_1$ describe the fits using bolometric luminosities. Because the center excision effectively eliminates the intrinsic scatter in the luminosity--mass relation (residual systematic scatter is bounded at the few per cent level in both cases), the  data can be fit directly, without using the full apparatus discussed in \secref{sec:analysis} and \cosmopaper{}. These constraints are marginalized over the range of cosmological parameter values allowed by the combination of XLF and other cosmological data.}
  \label{tab:ceslopes}
  \begin{tabular}{l@{\hspace{2mm}}c@{\hspace{2mm}}cc}
    \hline
    Parameter & & Standard & Extended \\
    \hline
    $\mysub{\ell}{ce}$--$m$ normalization      & $\gamma_0^{\ell m}$          & $0.71 \pm 0.06$ & $0.71 \pm 0.06$\vspace{0.5mm} \\
    $\mysub{\ell}{ce}$--$m$ slope              & $\gamma_1^{\ell m}$          & $1.02 \pm 0.05$ & $1.03 \pm 0.04$\vspace{0.5mm} \\
    $t$--$\mysub{\ell}{ce}$ normalization      & $\gamma_0^{t\ell}$         & $0.54 \pm 0.05$ & $0.55 \pm 0.05$\vspace{0.5mm} \\
    $t$--$\mysub{\ell}{ce}$ slope              & $\gamma_1^{t\ell}$         & $0.48 \pm 0.04$ & $0.47 \pm 0.04$\tablespace \\
    $\mysub{\ell}{bol,ce}$--$m$ normalization  & $\tilde{\gamma}_0^{\ell m}$  & $1.16 \pm 0.08$ & $1.16 \pm 0.08$\vspace{0.5mm} \\
    $\mysub{\ell}{bol,ce}$--$m$ slope          & $\tilde{\gamma}_1^{\ell m}$  & $1.31 \pm 0.05$ & $1.30 \pm 0.05$\vspace{0.5mm} \\
    $t$--$\mysub{\ell}{bol,ce}$ normalization  & $\tilde{\gamma}_0^{t\ell}$ & $0.45 \pm 0.06$ & $0.45 \pm 0.06$\vspace{0.5mm} \\
    $t$--$\mysub{\ell}{bol,ce}$ slope          & $\tilde{\gamma}_1^{t\ell}$ & $0.37 \pm 0.03$ & $0.38 \pm 0.03$ \\
    \hline
  \end{tabular}
\end{table}

The fit was marginalized over the cosmological parameter values allowed by the combination of XLF and other cosmological data (\cosmopaper{}). Note that, unlike elsewhere in this paper, we do not simultaneously fit for cosmological parameters here, but only marginalize the results over the allowed set of cosmological models; hence the fits in this section directly use only the clusters shown in \figref{fig:Lcice}, not the additional clusters that were detected in the RASS but not followed up.

From the standard data set (BCS, REFLEX and Bright MACS), we find normalization $\gamma_0^{\ell m}=0.71 \pm 0.06$ and slope $\gamma_1^{\ell m}=1.02 \pm 0.05$, with an upper bound of 0.015 on the intrinsic scatter (68.3 per cent confidence), for the self-similarly evolving ($\gamma_2^{\ell m}=0$) case. The extended data set, using the high-redshift MACS clusters and 400d clusters, yields virtually identical constraints. Similar results are obtained with the other simple fitting procedures discussed in \secref{sec:minmodres}. These results, and those for the corresponding bolometric luminosity--mass relation, are shown in \tabref{tab:ceslopes}. Note that, unlike the procedure in \secref{sec:minmodres}, fitting the bolometric relation in this case does not require the use of a temperature--mass relation; we directly fit the masses and bolometric luminosities derived from the follow-up X-ray observations.

\begin{table*}
  \centering
  \caption{Details of the follow-up observations of high-redshift ($z>0.5$) MACS clusters and 400d clusters (see caption for \tabref{tab:bcsobs}).}
  \begin{tabular}{|lccc@{~}c@{~}clcr@{.}l|}
    \hline
    Name & RA (J2000) & Dec (J2000) & \multicolumn{3}{c}{Date} & Detector & Mode & \multicolumn{2}{c}{Exposure (ks)} \\
    \hline
MACS J2214.9$-$1359  &  22  14  57.3  &  $-$14  00  11  &  2002  &  Dec  &  22  &  ACIS$-$I  &  VFAINT  &  \hspace{5ex}14  &  6  \\
                     &                &                 &  2003  &  Nov  &  17  &  ACIS$-$I  &  VFAINT  &  14              &  7  \\
MACS J0911.2+1746    &  09  11  10.9  &  +17    46  31  &  2003  &  Feb  &  23  &  ACIS$-$I  &  VFAINT  &  17              &  4  \\
                     &                &                 &  2004  &  Mar  &  08  &  ACIS$-$I  &  VFAINT  &  23              &  8  \\
MACS J0257.1$-$2325  &  02  57  9.1   &  $-$23  26  04  &  2000  &  Oct  &  03  &  ACIS$-$I  &  FAINT   &  19              &  6  \\
                     &                &                 &  2003  &  Aug  &  23  &  ACIS$-$I  &  VFAINT  &  18              &  2  \\
MACS J0454.1$-$0300  &  04  54  11.4  &  $-$03  00  51  &  2000  &  Jan  &  14  &  ACIS$-$I  &  VFAINT  &  13              &  8  \\
                     &                &                 &  2000  &  Oct  &  08  &  ACIS$-$S  &  FAINT   &  39              &  9  \\
MACS J1423.8+2404    &  14  23  47.9  &  +24    04  43  &  2001  &  Jun  &  01  &  ACIS$-$I  &  VFAINT  &  18              &  5  \\
                     &                &                 &  2003  &  Aug  &  18  &  ACIS$-$S  &  VFAINT  &  114             &  5  \\
MACS J1149.5+2223    &  11  49  35.4  &  +22    24  04  &  2001  &  Jun  &  01  &  ACIS$-$I  &  VFAINT  &  17              &  0  \\
                     &                &                 &  2003  &  Feb  &  07  &  ACIS$-$I  &  VFAINT  &  17              &  7  \\
MACS J0717.5+3745    &  07  17  32.1  &  +37    45  21  &  2001  &  Jan  &  29  &  ACIS$-$I  &  FAINT   &  16              &  5  \\
                     &                &                 &  2003  &  Jan  &  08  &  ACIS$-$I  &  VFAINT  &  55              &  1  \\
MACS J0018.5+1626    &  00  18  33.4  &  +16    26  13  &  2000  &  Aug  &  18  &  ACIS$-$I  &  VFAINT  &  66              &  9  \\
MACS J0025.4$-$1222  &  00  25  29.9  &  $-$12  22  45  &  2002  &  Nov  &  11  &  ACIS$-$I  &  VFAINT  &  15              &  7  \\
                     &                &                 &  2004  &  Aug  &  09  &  ACIS$-$I  &  VFAINT  &  24              &  4  \\
MACS J2129.4$-$0741  &  21  29  25.7  &  $-$07  41  31  &  2002  &  Dec  &  23  &  ACIS$-$I  &  VFAINT  &  10              &  5  \\
                     &                &                 &  2003  &  Oct  &  18  &  ACIS$-$I  &  VFAINT  &  19              &  3  \\
MACS J0647.7+7015    &  06  47  49.7  &  +70    14  56  &  2002  &  Oct  &  31  &  ACIS$-$I  &  VFAINT  &  16              &  7  \\
                     &                &                 &  2003  &  Oct  &  07  &  ACIS$-$I  &  VFAINT  &  19              &  9  \\
MACS J0744.8+3927    &  07  44  52.3  &  +39    27  27  &  2001  &  Nov  &  12  &  ACIS$-$I  &  VFAINT  &  19              &  7  \\
                     &                &                 &  2003  &  Jan  &  04  &  ACIS$-$I  &  VFAINT  &  15              &  9  \\
                     &                &                 &  2004  &  Dec  &  03  &  ACIS$-$I  &  VFAINT  &  49              &  5  \\
400d J0302.3$-$0423  &  03  02  21.1  &  $-$04  23  24  &  2005  &  Dec  &  07  &  ACIS$-$I  &  VFAINT  &  9              &  0  \\
400d J1212.3+2733    &  12  12  18.4  &  +27    33  02  &  2005  &  Mar  &  17  &  ACIS$-$I  &  VFAINT  &  12             &  5  \\
400d J0318.5$-$0302  &  03  18  33.3  &  $-$03  02  58  &  2005  &  Mar  &  15  &  ACIS$-$I  &  VFAINT  &  10             &  7  \\
400d J0809.6+2811    &  08  09  41.0  &  +28    12  01  &  2004  &  Nov  &  30  &  ACIS$-$I  &  VFAINT  &  5              &  3  \\
400d J1416.4+4446    &  14  16  28.1  &  +44    46  43  &  1999  &  Dec  &  02  &  ACIS$-$I  &  VFAINT  &  25             &  0  \\
400d J1701.3+6414    &  17  01  23.1  &  +64    14  08  &  2000  &  Oct  &  31  &  ACIS$-$I  &  VFAINT  &  37             &  1  \\
400d J0355.9$-$3741  &  03  55  59.4  &  $-$37  41  46  &  2006  &  Jan  &  12  &  ACIS$-$I  &  VFAINT  &  22             &  3  \\
400d J1002.1+6858    &  10  02  09.0  &  +68    58  36  &  2005  &  Jan  &  05  &  ACIS$-$I  &  VFAINT  &  17             &  9  \\
400d J0030.5+2618    &  00  30  33.8  &  +26    18  09  &  2005  &  May  &  28  &  ACIS$-$I  &  VFAINT  &  11             &  8  \\
400d J1524.6+0957    &  15  24  39.8  &  +09    57  46  &  2002  &  Apr  &  01  &  ACIS$-$I  &  VFAINT  &  40             &  7  \\
400d J1357.3+6232    &  13  57  17.6  &  +62    32  51  &  2006  &  Jan  &  24  &  ACIS$-$I  &  VFAINT  &  21             &  0  \\
                     &                &                 &  2006  &  Jan  &  29  &  ACIS$-$I  &  VFAINT  &  15             &  8  \\
400d J1120.9+2326    &  11  20  57.5  &  +23    26  29  &  2001  &  Apr  &  23  &  ACIS$-$I  &  VFAINT  &  55             &  0  \\
400d J0956.0+4107    &  09  56  03.3  &  +41    07  08  &  2003  &  Dec  &  30  &  ACIS$-$I  &  VFAINT  &  15             &  0  \\
                     &                &                 &  2005  &  Jan  &  28  &  ACIS$-$I  &  VFAINT  &  35             &  6  \\
400d J0328.6$-$2140  &  03  28  36.2  &  $-$21  40  22  &  2005  &  Mar  &  15  &  ACIS$-$I  &  VFAINT  &  29             &  4  \\
                     &                &                 &  2005  &  Mar  &  18  &  ACIS$-$I  &  VFAINT  &  3              &  7  \\
400d J1120.1+4318    &  11  20  06.9  &  +43    18  05  &  2005  &  Jan  &  11  &  ACIS$-$I  &  VFAINT  &  15             &  2  \\
400d J1334.3+5031    &  13  34  20.1  &  +50    31  01  &  2005  &  Aug  &  05  &  ACIS$-$I  &  VFAINT  &  7              &  0  \\
400d J0542.8$-$4100  &  05  42  50.1  &  $-$41  00  04  &  2000  &  Jul  &  26  &  ACIS$-$I  &  FAINT   &  41             &  1  \\
400d J1202.2+5751    &  12  02  17.8  &  +57    51  54  &  2005  &  Sep  &  02  &  ACIS$-$I  &  VFAINT  &  45             &  5  \\
400d J0405.4$-$4100  &  04  05  24.5  &  $-$41  00  19  &  2005  &  Oct  &  27  &  ACIS$-$I  &  VFAINT  &  6              &  6  \\
                     &                &                 &  2006  &  May  &  19  &  ACIS$-$I  &  VFAINT  &  57             &  7  \\
400d J1221.3+4918    &  12  21  26.1  &  +49    18  31  &  2001  &  Aug  &  05  &  ACIS$-$I  &  VFAINT  &  62             &  7  \\
400d J0152.7$-$1358  &  01  52  41.1  &  $-$13  58  07  &  2000  &  Sep  &  08  &  ACIS$-$I  &  FAINT   &  29             &  2  \\
400d J1226.9+3332    &  12  26  57.9  &  +33    32  49  &  2000  &  Jul  &  31  &  ACIS$-$S  &  VFAINT  &  9              &  8  \\
                     &                &                 &  2003  &  Jan  &  27  &  ACIS$-$I  &  VFAINT  &  26             &  1  \\
                     &                &                 &  2004  &  Aug  &  07  &  ACIS$-$I  &  VFAINT  &  25             &  3  \\
    \hline
  \end{tabular}
  \label{tab:moreobs}
\end{table*}

\begin{table*}
  \centering
  \caption{Redshifts and derived properties of high-redshift ($z>0.5$) MACS clusters and 400d clusters from follow-up observations (see caption for \tabref{tab:mltbcs}). Note that these data are used only in \secref{sec:mlce}.}
  \begin{tabular}{|llr@{$~\pm~$}lr@{$~\pm~$}lr@{$~\pm~$}lr@{$~\pm~$}l|}
    \hline
    Name & \multicolumn{1}{c}{$z$} & \multicolumn{2}{c}{$r_{500}$} & \multicolumn{2}{c}{$\Mgas{}_{,500}$} & \multicolumn{2}{c}{$M_{500}$} & \multicolumn{2}{c}{$\mysub{L}{500,ce}$}\\
    & & \multicolumn{2}{c}{(Mpc)} & \multicolumn{2}{c}{$(10^{14}\Msun)$} & \multicolumn{2}{c}{$(10^{14}\Msun)$} & \multicolumn{2}{c}{$(10^{44}\erg\second^{-1})$} \\
    \hline
MACS J2214.9$-$1359  &  0.5027  &  $1.39$  &  $0.08$  &  $1.51$  &  $0.22$  &  $13.2$  &  $2.3$  &  $8.28$   &  $0.37$  \\
MACS J0911.2+1746    &  0.5049  &  $1.22$  &  $0.06$  &  $1.03$  &  $0.11$  &  $9.0$   &  $1.2$  &  $5.41$   &  $0.25$  \\
MACS J0257.1$-$2325  &  0.5049  &  $1.20$  &  $0.06$  &  $0.97$  &  $0.13$  &  $8.5$   &  $1.3$  &  $6.59$   &  $0.30$  \\
MACS J0454.1$-$0300  &  0.5377  &  $1.31$  &  $0.06$  &  $1.31$  &  $0.13$  &  $11.5$  &  $1.5$  &  $9.50$   &  $0.41$  \\
MACS J1423.8+2404    &  0.5431  &  $1.09$  &  $0.05$  &  $0.76$  &  $0.08$  &  $6.6$   &  $0.9$  &  $4.64$   &  $0.20$  \\
MACS J1149.5+2223    &  0.5444  &  $1.53$  &  $0.08$  &  $2.13$  &  $0.28$  &  $18.7$  &  $3.0$  &  $12.61$  &  $0.52$  \\
MACS J0717.5+3745    &  0.5458  &  $1.69$  &  $0.06$  &  $2.85$  &  $0.24$  &  $24.9$  &  $2.7$  &  $18.03$  &  $0.71$  \\
MACS J0018.5+1626    &  0.5456  &  $1.47$  &  $0.08$  &  $1.88$  &  $0.24$  &  $16.5$  &  $2.5$  &  $11.84$  &  $0.67$  \\
MACS J0025.4$-$1222  &  0.5843  &  $1.12$  &  $0.04$  &  $0.87$  &  $0.08$  &  $7.6$   &  $0.9$  &  $6.81$   &  $0.31$  \\
MACS J2129.4$-$0741  &  0.5889  &  $1.25$  &  $0.06$  &  $1.22$  &  $0.14$  &  $10.6$  &  $1.4$  &  $7.88$   &  $0.39$  \\
MACS J0647.7+7015    &  0.5907  &  $1.26$  &  $0.06$  &  $1.25$  &  $0.15$  &  $10.9$  &  $1.6$  &  $8.07$   &  $0.37$  \\
MACS J0744.8+3927    &  0.6976  &  $1.26$  &  $0.06$  &  $1.43$  &  $0.15$  &  $12.5$  &  $1.6$  &  $9.07$   &  $0.34$  \\
400d J0302.3$-$0423  &  0.350  &  $0.99$  &  $0.07$  &  $0.46$  &  $0.09$  &  $4.0$   &  $0.8$  &  $2.52$  &  $0.24$  \\
400d J1212.3+2733    &  0.353  &  $1.35$  &  $0.10$  &  $1.18$  &  $0.22$  &  $10.3$  &  $2.1$  &  $5.20$  &  $0.35$  \\
400d J0318.5$-$0302  &  0.370  &  $0.85$  &  $0.11$  &  $0.30$  &  $0.07$  &  $2.6$   &  $0.7$  &  $1.96$  &  $0.17$  \\
400d J0809.6+2811    &  0.399  &  $1.07$  &  $0.11$  &  $0.62$  &  $0.15$  &  $5.4$   &  $1.4$  &  $3.23$  &  $0.30$  \\
400d J1416.4+4446    &  0.400  &  $0.83$  &  $0.05$  &  $0.29$  &  $0.05$  &  $2.5$   &  $0.5$  &  $1.71$  &  $0.12$  \\
400d J1701.3+6414    &  0.453  &  $0.90$  &  $0.06$  &  $0.38$  &  $0.07$  &  $3.4$   &  $0.7$  &  $2.31$  &  $0.14$  \\
400d J0355.9$-$3741  &  0.473  &  $0.83$  &  $0.05$  &  $0.32$  &  $0.06$  &  $2.8$   &  $0.5$  &  $1.65$  &  $0.15$  \\
400d J0030.5+2618    &  0.500  &  $0.82$  &  $0.05$  &  $0.31$  &  $0.06$  &  $2.7$   &  $0.5$  &  $1.96$  &  $0.21$  \\
400d J1002.1+6858    &  0.500  &  $0.93$  &  $0.07$  &  $0.46$  &  $0.09$  &  $4.0$   &  $0.8$  &  $2.16$  &  $0.22$  \\
400d J1524.6+0957    &  0.516  &  $0.94$  &  $0.07$  &  $0.49$  &  $0.08$  &  $4.2$   &  $0.8$  &  $2.81$  &  $0.20$  \\
400d J1357.3+6232    &  0.525  &  $0.81$  &  $0.05$  &  $0.31$  &  $0.05$  &  $2.7$   &  $0.5$  &  $1.93$  &  $0.12$  \\
400d J1120.9+2326    &  0.562  &  $0.85$  &  $0.16$  &  $0.39$  &  $0.10$  &  $3.4$   &  $0.9$  &  $2.27$  &  $0.16$  \\
400d J0956.0+4107    &  0.587  &  $0.78$  &  $0.06$  &  $0.30$  &  $0.05$  &  $2.6$   &  $0.4$  &  $2.15$  &  $0.12$  \\
400d J0328.6$-$2140  &  0.590  &  $0.86$  &  $0.07$  &  $0.40$  &  $0.10$  &  $3.5$   &  $0.9$  &  $2.44$  &  $0.19$  \\
400d J1120.1+4318    &  0.600  &  $0.98$  &  $0.07$  &  $0.61$  &  $0.12$  &  $5.3$   &  $1.1$  &  $4.61$  &  $0.41$  \\
400d J1334.3+5031    &  0.620  &  $0.85$  &  $0.10$  &  $0.41$  &  $0.13$  &  $3.6$   &  $1.2$  &  $2.27$  &  $0.63$  \\
400d J0542.8$-$4100  &  0.642  &  $0.99$  &  $0.08$  &  $0.64$  &  $0.14$  &  $5.6$   &  $1.3$  &  $3.64$  &  $0.26$  \\
400d J1202.2+5751    &  0.677  &  $0.80$  &  $0.05$  &  $0.36$  &  $0.07$  &  $3.2$   &  $0.7$  &  $2.56$  &  $0.23$  \\
400d J0405.4$-$4100  &  0.686  &  $0.72$  &  $0.04$  &  $0.26$  &  $0.04$  &  $2.3$   &  $0.4$  &  $2.36$  &  $0.16$  \\
400d J1221.3+4918    &  0.700  &  $0.97$  &  $0.05$  &  $0.65$  &  $0.09$  &  $5.7$   &  $0.9$  &  $4.27$  &  $0.27$  \\
400d J0152.7$-$1358  &  0.833  &  $0.97$  &  $0.26$  &  $0.89$  &  $0.33$  &  $7.8$   &  $3.0$  &  $6.66$  &  $0.51$  \\
400d J1226.9+3332    &  0.888  &  $1.00$  &  $0.05$  &  $0.89$  &  $0.10$  &  $7.8$   &  $1.1$  &  $7.66$  &  $0.33$  \\
    \hline
  \end{tabular}
  \label{tab:mltmore}
\end{table*}

We stress that the limit on the intrinsic scatter above must be treated with caution. The luminosities and masses determined from follow-up observations incorporate various systematic allowances, most importantly those accounting for instrument calibration and the cluster-to-cluster variation of \fgas{} (\secref{sec:deproj}). The precise definition of these allowances has very little impact on the mean relation derived here, or on the results of \secref{sec:results}; however, because of its small size, the scatter in the center-excised luminosity--mass relation is sensitive to these allowances. To estimate the effect that the systematic allowances have on the determination of the intrinsic scatter, we repeat the fit using statistical error bars only (as shown in \figref{fig:Lcice}). In this case, we find an intrinsic scatter of $0.025 \pm 0.004$, or $\sim 6$ per cent. We conclude that, while the definition of a formal confidence interval for the center-excised intrinsic scatter is problematic, the scatter is likely to be very small ($<10$ per cent).

The small size of this scatter in the self-similar case strongly suggests that departures from self-similar evolution in the center-excised relation are also small. We test this possibility by fitting \eqnref{eq:MLce} with the additional parameter $\gamma_2^{\ell m}$ free, finding $\gamma_2^{\ell m}=-0.12^{+0.39}_{-0.54}$ from the standard data set and $\gamma_2^{\ell m}=-0.22 \pm 0.25$ from the extended data, both consistent with zero. The latter corresponds to a 10 per cent constraint on excess evolution in the normalization of the relation within $z=0.5$ (or 16 per cent within $z=0.9$). The constraints on the slope and scatter are similar to those in \tabref{tab:ceslopes}.

The existence of such a simply evolving, low-scatter proxy for total mass is exciting in the context of future X-ray cluster searches. Since the center-excised luminosity at soft energies is derived from the count rate with little temperature dependence, it can be measured precisely without the need for high-quality spectral information. This is in contrast to other low-scatter mass proxies such as the average temperature ($kT$) or thermal energy ($Y_X=\Mgas \mysub{kT}{ce}$), which are generally limited by the precision of temperature measurements. Although the upcoming eROSITA\footnote{\url{http://www.mpe.mpg.de/projects.html\#erosita}} survey is unlikely to have sufficient spatial resolution to directly measure center-excised fluxes, a mission along the lines of the proposed Wide-Field X-ray Telescope \citep[WFXT,][]{Murray08} could potentially take advantage of this tight relation to construct an effectively mass-limited cluster sample directly from an X-ray center-excised flux-limited survey.

If the scatter between mass and center-excised luminosity is negligible, the dependence of the center-excised temperature--luminosity relation on the mass function (\eqnref{eq:LTrel}) vanishes. The parameters of the nominal, center-excised temperature--luminosity relation,
\begin{equation}
  \label{eq:LceT}
  \expectation{t(\mysub{\ell}{ce})} = \gamma_0^{t\ell} + \gamma_1^{t\ell} \mysub{\ell}{ce},
\end{equation}
can then be estimated simply as (\tabref{tab:ceslopes})
\begin{eqnarray}
  \label{eq:LTsimple}
  \gamma_0^{t\ell} &=& \beta_0^{tm} - \beta_1^{tm} \frac{\gamma_0^{\ell m}}{\gamma_1^{\ell m}} , \\
  \gamma_1^{t\ell} &=& \frac{\beta_1^{tm}}{\gamma_1^{\ell m}}. \nonumber
\end{eqnarray}
The intrinsic scatter is identical to that of the temperature--mass relation.

\section{Astrophysical implications} \label{sec:astroimp}

The virial theorem makes specific predictions for the slopes of the scaling relations linking cluster mass, bolometric luminosity and mass-weighted temperature; namely
\begin{eqnarray}
  \label{eq:virialpr}
  \frac{\mysub{L}{bol}}{E(z)} & \propto & \left[ E(z)M \right]^{4/3}, \nonumber \\
  \mysub{kT}{mw} & \propto & \left[ E(z)M \right]^{2/3}, \\
  \mysub{kT}{mw} & \propto & \left[ \frac{\mysub{L}{bol}}{E(z)} \right]^{1/2}, \nonumber
\end{eqnarray}
for quantities within a fixed critical-overdensity radius. Compared to this prediction, our result for the (center-included) bolometric luminosity--mass slope, $\tilde{\beta}_1^{\ell m}=1.63 \pm 0.06$, is significantly steeper, and our constraint on the temperature--mass slope, $\beta_1^{tm}=0.49 \pm 0.04$, is significantly shallower. These results are consistent with a picture in which the intracluster medium has been subject to excess heating;\footnote{However, as noted below, the fact that emission-weighted rather than mass-weighted temperatures are measured complicates the physical interpretation of the temperature--mass slope. In this work, the luminosity--mass relation is preferred for studying the effects of excess heating of the ICM.} energy injection heats the gas, raising the temperature and decreasing the density (hence the luminosity), with the effect being more prominent in lower mass systems. Condensation of the coldest gas in cluster cores, leaving behind hotter, less dense gas, may also play a role, although some additional heating is still required in order to account for the typically modest star formation rates observed in cool cores \citep[e.g.][]{Voit01,Peterson06,McNamara07}. Importantly, the consistency of our data with a simple, self-similarly evolving power-law model indicates that the effects of the excess heat on the global luminosity and temperature have not evolved strongly since redshift 0.5.

Interestingly, when the central $0.15r_{500}$ is excluded from the luminosity measurements, the slope of the bolometric luminosity--mass relation, $\tilde{\gamma}_1^{\ell m}=1.30 \pm 0.05$, is close to the virial prediction. This observation suggests that the excess heating is limited to the central regions of clusters, arguing in favor of local heating mechanisms, such as AGN feedback, and against mechanisms that would heat the ICM more uniformly, such as preheating. The fact that the center-excised temperature--mass slope does not agree with the virial prediction does not necessarily change this picture, since that prediction applies to the mass-weighted temperature rather than the emission-weighted temperatures reported here.\footnote{Unfortunately, it is not currently feasible to measure mass-weighted temperatures within $r_{500}$ due to systematic uncertainties in the background modeling.} Note also that the emission-weighted temperature is more influenced by gas at small radii, i.e. closer to the source of excess heating, where the signal-to-noise is much higher than in the outskirts. The small intrinsic scatter of the center-excised luminosity--mass (particularly) and temperature--mass relations over a wide range in redshift, including terms for self-similar evolution as we have, further reinforces the notion that the self-similar model closely describes the ICM outside of cluster centers.

\section{Conclusion} \label{sec:conclusion}

We have presented constraints on the X-ray scaling relations of galaxy clusters using a new analysis method, detailed in \cosmopaper{}, that self-consistently accounts for all observational biases and degeneracies with cosmological parameters. The data are drawn from three wide area, X-ray flux-limited cluster samples based on the RASS, span the redshift range $0<z<0.5$, and satisfy the luminosity threshold $L>2.5\E{44}h_{70}^{-2}\erg\second^{-1}$. Masses, X-ray luminosities, and average ICM temperatures at $r_{500}$ were derived from follow-up X-ray observations of 94 of the 238 flux-selected clusters, and used to constrain joint luminosity--temperature--mass scaling relations, marginalizing over a \LCDM{} cosmological background as well as conservative systematic allowances. Our analysis for the first time incorporates the cluster mass function and sample selection function in a fully self-consistent and statistically rigorous way.

The data are consistent with a relatively simple picture in which the nominal temperature--mass and luminosity--mass relations in the observed mass range are described by power laws that evolve self similarly. The scatter in luminosity and temperature at fixed mass is adequately described by a simple, bivariate log-normal distribution. Departures from self-similar evolution, evolution in the scatter, and asymmetry in the scatter were considered, but the data do not require any of these additions to the model at a statistically significant level. Incorporating additional cosmological data to minimize the degeneracies with cosmological parameters, we find that the slope of the (center-included) luminosity--mass relation is $\beta_1^{\ell m}=1.34 \pm 0.05$ (corresponding to bolometric luminosity--mass slope $1.63 \pm 0.06$) while the slope of the (center-excised) temperature--mass relation is $\beta_1^{tm}=0.49 \pm 0.04$. The marginal scatters of the relations are $\sigma_{\ell m}=0.185 \pm 0.019$ and $\sigma_{tm}=0.055 \pm 0.008$ (68.3 per cent confidence intervals). The correlation coefficient of the luminosity and temperature scatters at fixed mass is relatively poorly constrained, but consistent with zero, $\rho_{\ell t m}=0.09 \pm 0.19$. The resulting temperature--luminosity relation has slope $\beta_1^{t\ell}=0.29 \pm 0.08$ and intrinsic scatter $\sigma_{t\ell}=0.08 \pm 0.02$. The steep slope of the bolometric luminosity--mass relation and shallow slope of the temperature--mass relation relative to the predictions of the virial theorem (respectively $4/3$ and $2/3$) are consistent with the now-standard picture of excess heating of the ICM. Moreover, the agreement of our data with self-similar evolution indicates that the heating mechanism was in operation before $z=0.5$, and that its effects on global cluster properties do not evolve strongly at redshifts $z<0.5$.

Previous observations that the scatter in the luminosity--mass relation is dominated by the emission in the cluster center ($r<0.15r_{500}$) are reinforced by our data. Here, we find three main, new results. First, the intrinsic scatter in the center-excised luminosity--mass relation is consistent with zero; we derive an upper bound at the $<10$ per cent level. Second, the center-excised, bolometric luminosity--mass slope, $\tilde{\gamma}_1^{\ell m}=1.30 \pm 0.05$, is in agreement with the virial prediction. Third, we find that the evolution of this relation to be consistent with the self-similar expectation, and bound departures from self similarity at the 16 per cent level over the range $z<0.9$ (10 per cent over $z<0.5$). Together, these results argue that the excess heating in clusters has been primarily limited to their centers, supporting local heating mechanisms such as AGN feedback over more global preheating.

The existence of a tight, simply evolving correlation between center-excised luminosity and mass is exciting for cosmological applications, since the center-excised luminosity may provide a simpler and cheaper (in terms of required exposure time) mass proxy than average temperature or thermal energy. Potentially, it could form the basis for the selection of an effectively mass-limited cluster sample from a future X-ray survey, given an instrument with sufficient spatial resolution.

In this paper, we have not presented any cosmological results, despite the fact that the analysis used here simultaneously constrains cosmological parameters. Those results, as well as constraints on more complex cosmological models, can be found in \cosmopaper{}. The scaling relation results presented here should provide a benchmark for comparison with cosmological simulations of the formation and evolution of massive clusters including baryonic physics.

Monte Carlo samples encoding the results of this paper and \cosmopaper{} will be made available for download on the web.\footnote{\url{http://www.stanford.edu/group/xoc/papers/xlf2009.html}}

\section*{Acknowledgments} \label{sec:acknowledgements}

We thank the reviewer for a very careful reading of the manuscript, Dale Kocevski for providing redshifts of REFLEX clusters, and Eli Rykoff for helpful discussions. We are also grateful to Glenn Morris, Stuart Marshall and the SLAC unix support team for technical support. Calculations were carried out using the KIPAC XOC and Orange compute clusters at the SLAC National Accelerator Laboratory and the SLAC Unix compute farm. We acknowledge support from the National Aeronautics and Space Administration (NASA) through LTSA grant NAG5-8253, and though Chandra Award Numbers DD5-6031X, GO2-3168X, GO2-3157X, GO3-4164X, GO3-4157X, GO5-6133, GO7-8125X and GO8-9118X, issued by the Chandra X-ray Observatory Center, which is operated by the Smithsonian Astrophysical Observatory for and on behalf of NASA under contract NAS8-03060. This work was supported in part by the U.S. Department of Energy under contract number DE-AC02-76SF00515. AM was supported by a William~R. and Sara Hart Kimball Stanford Graduate Fellowship.

\vspace{-5mm}

\bibliographystyle{mnras}
\def \aap {A\&A} % alternative A&A code
\def \statisci {Statis. Sci.}
\def \physrep {Phys. Rep.}
\def \pre {Phys.\ Rev.\ E}
\def \sjos {Scand. J. Statis.} % Scandinavian Journal of Statistics
\def \jrssb {J. Roy. Statist. Soc. B} % Journal of the Royal Statistical Society. Series B (Statistical Methodology)
\def \pan {Phys. Atom. Nucl.} % Physics of Atomic Nuclei
\def \epja {Eur. Phys. J. A} % European Physical Journal A
\def \epjc {Eur. Phys. J. C} % European Physical Journal C
\def \jcap {J. Cosmology Astropart. Phys.} % Journal of Cosmology and Astro-Particle Physics
\def \ijmpd {Int.\ J.\ Mod.\ Phys.\ D}
\def \araa {ARA\&A}
\def \aj {AJ}
\def \apj {ApJ}
\def \apjl {ApJL}
\def \apjs {ApJS}
\def \mnras {MNRAS}
\def \nat {Nat}
\def \pasj {PASJ}
\def \gca {Geochim.\ Cosmochim.\ Acta}
\def \npa {Nucl.\ Phys.\ A}
\def \plb {Phys.\ Lett.\ B}
\def \prc {Phys.\ Rev.\ C}
\def \prd {Phys.\ Rev.\ D}
\def \prl {Phys.\ Rev.\ Lett.}

\appendix

\section{The effects of selection biases on the scaling relations} \label{sec:pedbias}

\begin{figure*}
  \centering
  \includegraphics[scale=0.85]{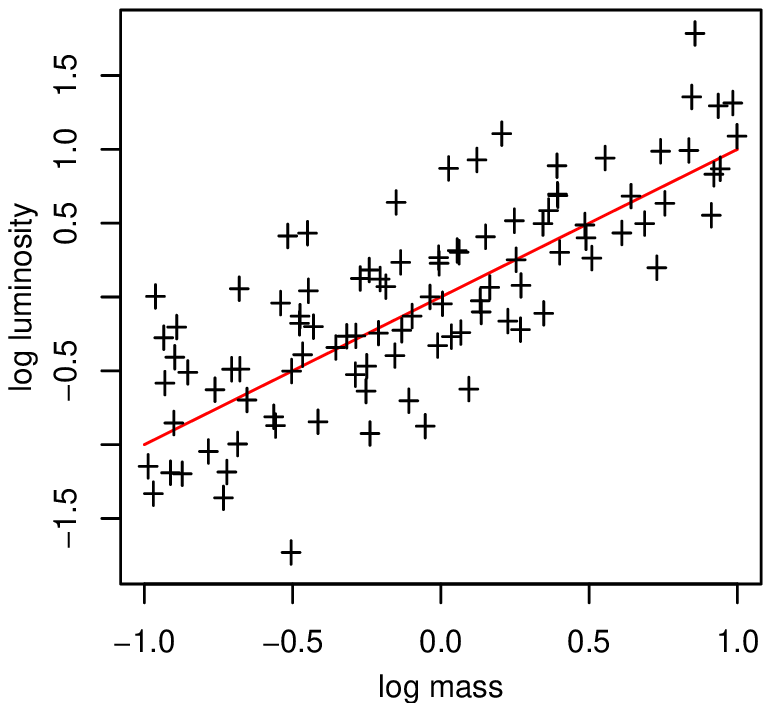}
  \hspace{0.5cm}
  \includegraphics[scale=0.85]{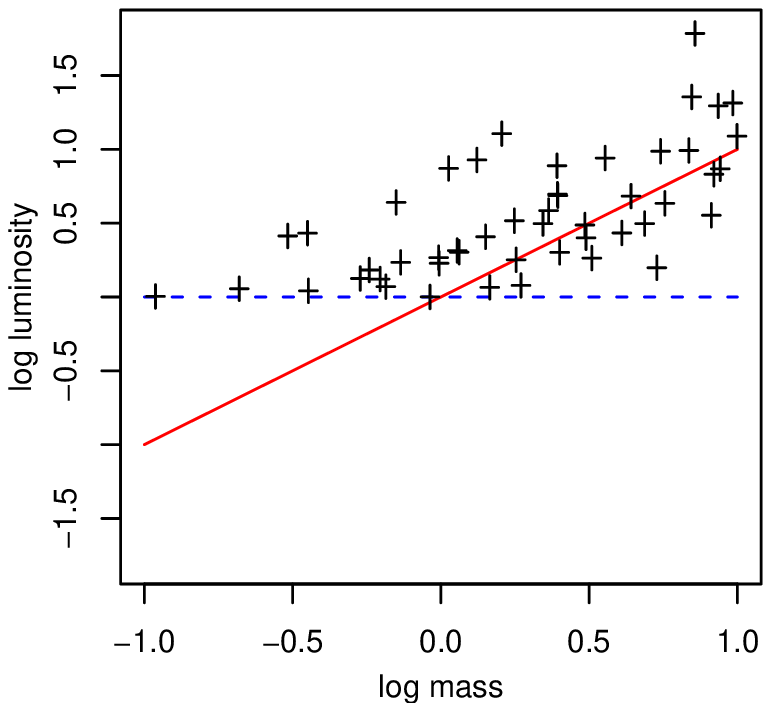}
  \vspace{0.cm} \\
  \includegraphics[scale=0.85]{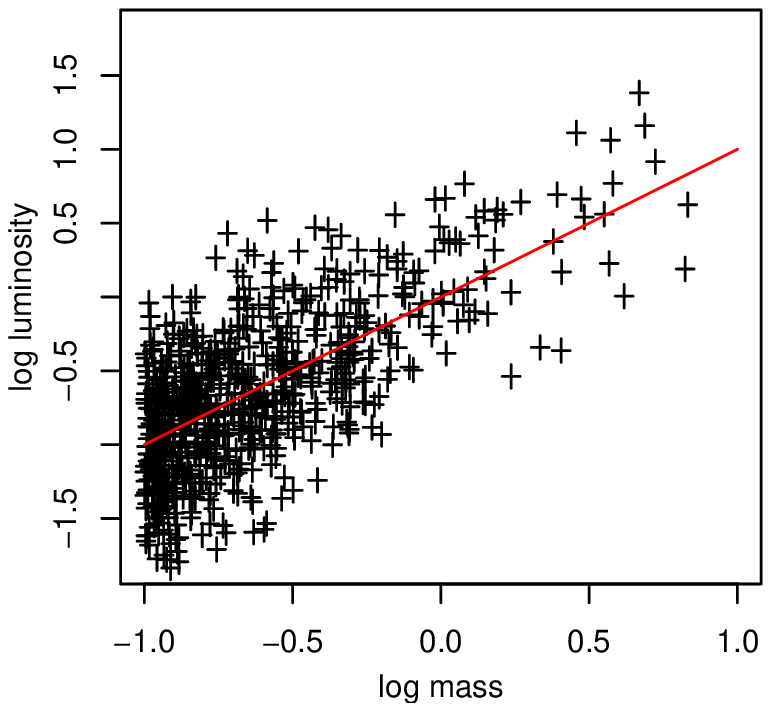}
  \hspace{0.5cm}
  \includegraphics[scale=0.85]{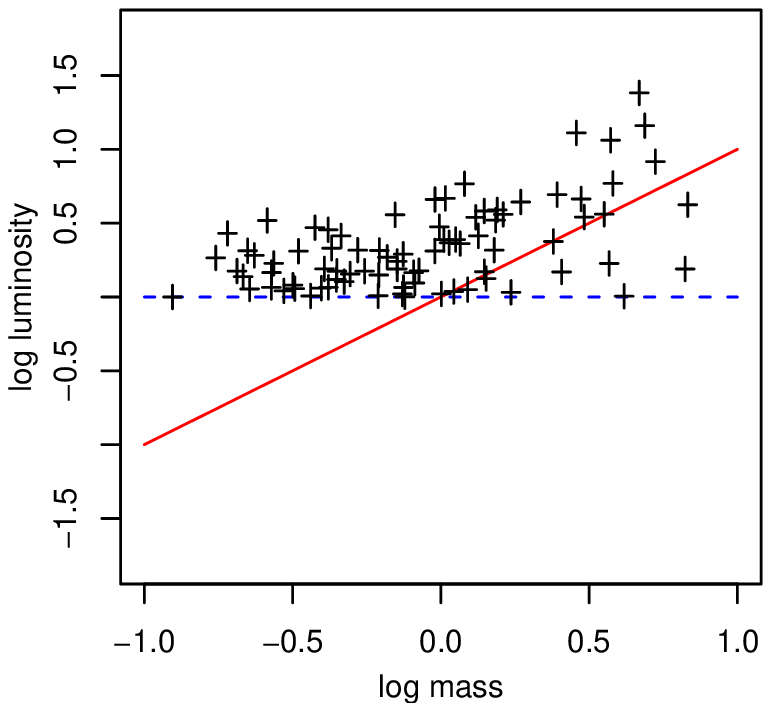}
  \caption{Fictitious cluster luminosity--mass relations (red lines) and simulated data (crosses) intended to illustrate the effect of Malmquist and Eddington biases on the scaling relation data. In the top panels, clusters are distributed uniformly in log-mass, whereas in the bottom panels the distribution of log-masses is exponential. The left-hand panels reflect the true distribution of all clusters in mass and luminosity, while the right-hand panels show only the simulated clusters with luminosities greater than a threshold value, indicated by the dashed, blue lines. The figure illustrates that both the sample selection function and the underlying mass function must be taken into account when fitting the scaling relations.}
  \label{fig:selbias}
\end{figure*}

Here we offer a pedagogical look at the effects of Malmquist and Eddington biases in the context of fitting the scaling relations. Consider the upper-left panel of \figref{fig:selbias}, which shows a fictitious set of cluster masses and luminosities (crosses) and the associated luminosity--mass relation (red line). The simulated clusters are distributed uniformly in log-mass, and scattered in luminosity about the nominal relation following a log-normal distribution. The simplest way to visualize the effect of Malmquist bias on this distribution of clusters is to consider a threshold luminosity below which no clusters can be detected; this results in the ``observed'' data set shown above the dashed, blue line in the upper-right panel of the figure (assuming complete follow-up and ignoring any additional scatter due to measurement errors). The distribution of observed clusters in the log-mass log-luminosity plane clearly does not follow the underlying relation indicated by the red line, and any simple method of fitting the data that does not account for the selection procedure (i.e. the luminosity threshold) will be biased by the low-mass clusters, which lie far from the mean relation.

This picture is further complicated by the presence of Eddington bias, which results from the fact that the number of clusters in the Universe decreases rapidly as a function of mass. Although usually thought of in terms of the total number of cluster detections, the steep slope of the mass function has the additional effect of exaggerating the biased nature of the scaling relation data. This can be seen in the lower panels of \figref{fig:selbias}, which show the true and observed distribution of cluster masses and luminosities relative to the true relation and the luminosity threshold when the distribution of cluster log-masses is exponential rather than uniform. Straightforwardly, the presence of many more low-mass than high-mass clusters results in a larger fraction of heavily biased (i.e. more luminous than average) data.

In practice, the effects of selection bias on our luminosity--mass data are less visually obvious (cf. \figref{fig:fitMLT}) because the data are drawn from three flux-limited samples, each with a different flux limit. In effect, the applicable threshold $\ell$ is a function of both redshift and the applicable flux limit, and therefore varies from cluster to cluster. Nevertheless, the data are certainly affected by selection bias, as reflected in the apparent offset between the data and the best-fitting relation from our analysis in \figref{fig:fitMLT}. We stress that re-measurement of cluster luminosities through follow-up observations does {\it not} eliminate the effects of the Malmquist and Eddington biases described above because most of the scatter in survey-measured luminosity at fixed mass is due to intrinsic scatter in the scaling relation, not measurement errors.

 Fitting methods that do not explicitly account for the distribution of covariates (i.e. that assume they are uniformly distributed) will naturally assign disproportionate importance to the lowest-mass systems, whose number is very sensitive to the details of the mass function in addition to the selection function. To properly compensate for this effect, it is thus crucial that an analysis incorporate information about both the mass function and sample selection function, as described in \cosmopaper{}. In practice, it is necessary to marginalize over cosmological parameters which affect the mass function, particularly $\sigma_8$.

\section{Predicting luminosity measurements} \label{sec:plzmi}

The distribution $P(\hat{\ell}'|\hat{z},\hat{m},I)$ describes the likelihood of a cluster having luminosity $\hat{\ell}'$ measured in follow-up observations, given that it was detected in the survey at redshift $\hat{z}$ and given that the mass measured in follow-up observations was $\hat{m}$ (see definitions in \eqnref{eq:MLTdefs}). In practice, the spectroscopic redshift measurement error is negligible, and we therefore set $P(\hat{z}|z) = \delta(z-\hat{z})$ (where $\delta$ is the Dirac delta function), i.e. $z=\hat{z}$. With this simplification, $P(\hat{\ell}'|\hat{z},\hat{m},I)$ can be written
\begin{eqnarray}
  P(\hat{\ell}'|\hat{z},\hat{m},I) & = & \int dm ~ P(m) \int d\ell \int dt ~ P(\ell,t|z,m) \nonumber \\
  & & \times ~ P(\hat{m},\hat{\ell}'|m,\ell,t) ~ P(I|z,\ell,t),
\end{eqnarray}
where $P(m)$ is proportional to the mass function at redshift $z$; $P(\ell,t|z,m)$ is the stochastic, bivariate scaling relation; $P(\hat{m},\hat{\ell}'|z,m,\ell,t)$ is the likelihood of measuring mass $\hat{m}$ and luminosity $\hat{\ell}'$ given true values $m$, $\ell$ and $t$; and $P(I|z,\ell,t)$ is the probability of detecting a cluster with redshift $z$, luminosity $\ell$ and temperature $t$ in the survey. This distribution is closely related to the quantity $\mysub{P}{det}$ defined in Section~4.1.2 of \cosmopaper{}, the primary difference being that the information provided by the survey flux measurement is not incorporated into in the expression above.

\bsp
\label{lastpage}
\end{document}